\documentstyle[12pt,epsf]{article}
\newdimen\rotdimen
\def\vspec#1{\special{ps:#1}}
\def\rotstart#1{\vspec{gsave currentpoint currentpoint translate
   #1 neg exch neg exch translate}}
\def\rotfinish{\vspec{currentpoint grestore moveto}}
\def\rotr#1{\rotdimen=\ht#1\advance\rotdimen by\dp#1%
   \hbox to\rotdimen{\hskip\ht#1\vbox to\wd#1{\rotstart{90 rotate}%
   \box#1\vss}\hss}\rotfinish}
\def\rotl#1{\rotdimen=\ht#1\advance\rotdimen by\dp#1%
   \hbox to\rotdimen{\vbox to\wd#1{\vskip\wd#1\rotstart{270 rotate}%
   \box#1\vss}\hss}\rotfinish}%
\def\rotu#1{\rotdimen=\ht#1\advance\rotdimen by\dp#1%
   \hbox to\wd#1{\hskip\wd#1\vbox to\rotdimen{\vskip\rotdimen
   \rotstart{-1 dup scale}\box#1\vss}\hss}\rotfinish}%
\def\rotf#1{\hbox to\wd#1{\hskip\wd#1\rotstart{-1 1 scale}%
   \box#1\hss}\rotfinish}%

\catcode`\@=11
\long\def\@makefntext#1{ 
\protect\noindent \hbox to 3.2pt {\hskip-.9pt  
$^{{\eightrm\@thefnmark}}$\hfil}#1\hfill} 

 \def\@makefnmark{\hbox to 0pt{$^{\@thefnmark}$\hss}}  
        
\def\ps@myheadings{\let\@mkboth\@gobbletwo
\def\@oddhead{\hbox{} 
\rightmark\hfil\eightrm\thepage}   
\def\@oddfoot{}\def\@evenhead{\eightrm\thepage\hfil 
\leftmark\hbox{}}\def\@evenfoot{}
\def\sectionmark##1{}\def\subsectionmark##1{}}

\oddsidemargin=\evensidemargin
\newcounter{sectionc}\newcounter{subsectionc}\newcounter{subsubsectionc}
\renewcommand{\section}[1] {\vspace{12pt}\addtocounter{sectionc}{1} 
\setcounter{subsectionc}{0}\setcounter{subsubsectionc}{0}\noindent 
        {\bf\thesectionc. #1}\par\vspace{5pt}}
\renewcommand{\subsection}[1] {\vspace{12pt}\addtocounter{subsectionc}{1} 
        \setcounter{subsubsectionc}{0}\noindent 
        {\bf\thesectionc.\thesubsectionc. {\kern1pt \bf\it #1}}\par\vspace{5pt}}
\renewcommand{\subsubsection}[1] {\vspace{12pt}\addtocounter{subsubsectionc}{1}
        \noindent{\thesectionc.\thesubsectionc.\thesubsubsectionc.
        {\kern1pt \it #1}}\par\vspace{5pt}}
\newcommand{\nonumsection}[1] {\vspace{12pt}\noindent{\bf #1}
        \par\vspace{5pt}}

\newcommand{\smalllineskip}{\baselineskip=12pt}

\def\eightcirc{
\begin{picture}(0,0)
\put(4.4,1.8){\circle{6.5}}
\end{picture}}
\def\eightcopyright{\eightcirc\kern2.7pt\hbox{\eightrm c}}

\newcounter{itemlistc}
\newcounter{romanlistc}
\newcounter{alphlistc}
\newcounter{arabiclistc}

\def\thebibliography#1{\nonumsection{\large \bf References}\list
  {[\arabic{enumi}]}{\settowidth\labelwidth{[#1]}\leftmargin\labelwidth
    \advance\leftmargin\labelsep
    \usecounter{enumi}}
    \def\newblock{\hskip .11em plus .33em minus .07em}
    \sloppy\clubpenalty4000\widowpenalty4000}

\def\pmb#1{\setbox0=\hbox{#1}
        \kern-.025em\copy0\kern-\wd0
        \kern.05em\copy0\kern-\wd0
        \kern-.025em\raise.0433em\box0}

\def\fnt#1#2{\footnotetext{\kern-.3em
        {$^{\mbox{\scriptsize #1}}$}{#2}}}

 1
 1
 1

\font\eightrm=cmr8
\font\eightit=cmti8
\font\sevenrm=cmr7

\def\qed{\hbox{${\vcenter{\vbox{                          
   \hrule height 0.4pt\hbox{\vrule width 0.4pt height 6pt
   \kern5pt\vrule width 0.4pt}\hrule height 0.4pt}}}$}}

\newcommand{\bea}{\begin{eqnarray}}
\newcommand{\eea}{\end{eqnarray}}
\newcommand{\be}{\begin{equation}}
\newcommand{\ee}{\end{equation}}

\def\shiftleft#1{#1\llap{#1\hskip 0.04em}}
\def\shiftdown#1{#1\llap{\lower.04ex\hbox{#1}}}
\def\thick#1{\shiftdown{\shiftleft{#1}}}
\def\b#1{\thick{\hbox{$#1$}}}
\begin{document}
%
%
%

\title{Electromagnetic Properties of the $\Delta (1232)$ }

\author{
A. J. Buchmann$^1$, E. Hern\'andez$^2$, and Amand Faessler$^1$   \\ \\
$^1$ Institut f\"ur Theoretische Physik, Universit\"at T\"ubingen\\
Auf der Morgenstelle 14, D-72076 T\"ubingen, Germany \\
$^2$ Grupo de Fisica Nuclear, Universidad de Salamanca  \\
E-37008 Salamanca, Spain  }
\bigskip
\maketitle
\bigskip
\begin{abstract}
We calculate the electromagnetic moments and radii of the 
$\Delta(1232)$ in the nonrelativistic quark model, including
two-body exchange currents. We show that two-body exchange currents 
lead to nonvanishing $\Delta$ and $N\to \Delta$ transition quadrupole moments 
even if the wave functions have no $D$-state admixture. The usual explanation
based on the single-quark transition model involves $D$-state admixtures
but no exchange currents.
We derive a parameter-free relation between the $N \to \Delta$ 
transition quadrupole moment and the neutron charge radius, namely 
$Q_{N \to \Delta}= {1\over \sqrt{2} } \, r^2_n$. 
Furthermore, we calculate the $M1$ and $E2$ 
amplitudes for the process $\gamma + N \to \Delta$. 
We find that  the $E2$ amplitude receives 
sizeable contributions from exchange currents. 
These are more important than the ones which result from $D$-state 
admixtures due to tensor forces between quarks if a reasonable quark core 
radius of about 0.6 fm is used. 
We obtain a ratio of $E2/M1=-3.5 \%$.
\end{abstract}
\vfill
\eject
%
%
%
%
\noindent
\section{\bf Introduction}
\par
\noindent
\par
Low-energy electromagnetic properties of baryons, such as charge radii,
magnetic moments and quadrupole moments, are very useful observables. 
In particular, $\Delta$ electromagnetic
properties provide valuable information on the quark-quark interaction 
that would otherwise be quite difficult to obtain. 
For example, while the nucleon wave function may contain a 
small D-state admixture and may therefore be deformed, angular momentum 
selection rules do not allow a spin 1/2 particle to have a nonzero 
quadrupole moment. However, as a spin 3/2 particle, the $\Delta$ can have an 
observable quadrupole moment. If this could be measured, it would provide 
additional constraints on the magnitude of the D-state admixture 
in baryon ground-state wave functions.
\par
In addition to the electromagnetic moments and radii of the $\Delta$,
the electromagnetic $\gamma + N \to \Delta$ transition form factors 
have received considerable attention during recent years. The reason is clear. 
While the magnetic and quadrupole moments of the $\Delta$ are very hard 
to measure, there are new high-precision pionproduction 
experiments with real and virtual photons in the $\Delta$-resonance 
region \cite{Eri94} which 
will provide accurate data on the electric quadrupole ($E2$) and 
magnetic dipole ($M1$) parts of the amplitude. 
These transition multipoles are sensitive to details of the quark dynamics. 
In particular, the $E2$ amplitude is crucial in getting a handle on the 
tensor forces between quarks and the related question of the deformation of 
the nucleon. 

\par
It has long been known that the reaction 
$ \gamma + p \to \Delta^+$ poses a problem for the additive quark model. 
The additive quark model predicts a relation between the
$M1$ transition moment and the proton magnetic moment \cite{Beg64}
\be 
\label{m1}
\mu_{p \to \Delta^+} 
= {2 \sqrt{2}\over 3} \mu_p.
\ee
Furthermore, it predicts that the $p\to \Delta^+$ quadrupole transition 
moment is exactly zero \cite{Mor65}
\be 
Q_{p \to \Delta^+} = 0.
\ee
Both results contradict experimental findings.
Eq.(\ref{m1}) gives $ \mu_{p \to \Delta^+} =2.63 \, \mu_N$
if the experimental proton magnetic moment is used. 
This is about $30 \%$ lower than the empirical 
value $3.5(2) \, \mu_N$\cite {Dal66}. Also, the empirical
quadrupole transition moment is small, but clearly nonzero \cite{Kal93}.
Various corrections to the simple nonrelativistic quark model (NRQM) 
results have been considered.
There are several works \cite{Kon80,Ger82,Dre84} which include  
$D$-state admixtures to the $N$ and $\Delta$ ground state 
resulting from one-gluon exchange induced tensor forces.
The inclusion of these $D$-states leads to a
nonvanishing $E2$ transition amplitude; however, the effect of $D$-state
admixtures on the $M1$ amplitude slightly increases the discrepancy 
between theory and experiment. 
Other authors calculate relativistic 
corrections to the single-quark current \cite{Clo90,War90,Cap92}. 
These corrections, although they are significant, are too small to account 
for the data.
The role of pions has been studied mainly in the 
framework of the bag model \cite{Ber88}, in effective Lagrangian
models \cite{Dav91,Chr92,Lag88,NBL90}, or in the Skyrme model 
\cite{Wir87,Aba96}.

\par
Recently, Robson \cite{Rob93} calculated the $A^{3/2}$ and $A^{1/2}$ 
helicity amplitudes for the $N \to \Delta$ transition including the 
pion pair exchange current but did not properly include the pionic current 
contribution. Also the $E2$ contribution to the helicity amplitudes was 
omitted in this first calculation. The contribution of pion tensor forces 
to the $E2$ transition was calculated in ref. \cite{Wey86}, but in this work 
exchange current corrections were omitted. 
Most calculations of electromagnetic properties in the constituent quark 
model (CQM)
have been performed in the so-called impulse approximation,
which assumes that the total electromagnetic current of the quarks
is given by a sum of {\it free} quark currents. 
However, a calculation
based on the impulse approximation is incomplete because it violates
current conservation.
Current conservation demands that the
total electromagnetic current operator of bound quarks
necessarily consist of two pieces: the one-body quark currents
and the two-body exchange currents associated with the
interactions responsible for quark binding.

\par
In two previous papers \cite{Buc91,Buc94} we have investigated the 
effect of two-body exchange currents on the charge and magnetic form
factors of the nucleon. 
We have shown that two-body gluon and pion
exchange currents are essential in simultaneously describing proton
and neutron charge radii and the positive parity excitation spectrum of the
nucleon with a single set of parameters.  
In particular, by including gluon and pion
exchange currents we were able to get a non-zero neutron charge radius of 
the right size. Using ground state wave functions we derived a 
relation between the neutron
charge radius, the $\Delta-N$ mass splitting, and the quark core radius
$b$, namely 
$r_n^2=-b^2(M_{\Delta}-M_N)/M_N$,
which clearly shows 
the underlying connection between the excitation spectrum 
of the nucleon and its electromagnetic 
properties.  Although this relation was derived with ground state wave 
functions, it remains approximately
valid even after the inclusion of configuration mixing \cite{Buc91}.
With respect to the magnetic moments, we have shown that 
exchange currents give individually large corrections but tend to
cancel each other globally \cite{Buc94,Wag95}, provided that the 
oscillator parameter is consistent with the one required to describe the
experimental neutron charge radius. Finally, we have noted that although 
the direct effect of pions on electromagnetic properties was not particularly 
large, their inclusion was essential for a satisfactory description of
the data. 

\par
In this work we extend our study of two-body exchange currents to the
electromagnetic moments and radii of the $\Delta$. Furthermore, we
calculate the magnetic and quadrupole $N-\Delta$ transition moments
and the corresponding helicity amplitudes $A_{3/2}$ and $A_{1/2}$ for
the photoexcitation of the $\Delta$. Instead of focusing on just
one observable (e.g. $E2/M1$ ), we simultaneously calculate a number of 
important low-energy observables including the 
magnetic and quadrupole moments, and the
charge and magnetic radii of {\it both} the nucleon and the $\Delta$, using 
a single set of parameters. In addition to the gluon, pion, and 
confinement exchange currents considered previously, we include the
$\sigma$-exchange current as suggested by chiral symmetry.
The aim of the present paper is to study the role 
of two-body exchange currents in $\Delta$ electromagnetic properties. 
In order to isolate and emphasize their contribution and to keep calculations
to a simple level we will take pure $L=0$, ground state,  harmonic oscillator,
orbital wave functions for both the nucleon and $\Delta$.

We show that two-body currents 
yield substantial corrections for the quadrupole moment of the
$\Delta$ and the quadrupole transition moment. We find that the
$C2$-amplitude is largely governed by the
spin-dependent two-body pieces in the charge density operator, which  
in the long wave-length limit also determine the $E2$-amplitude in
photo-pionproduction by application of Siegert's theorem.
This implies that the $E2$-transition to the $\Delta$ 
is presumably to a large extent a {\it two-quark spin-flip} 
transition.
\par
The paper is organized as follows. In sect. 2, we review the chiral constituent
quark model
which includes not only gluon, but also pion and sigma-meson exchange 
between constituent quarks.
Then, we list the two-body current operators connected with 
these quark-quark interactions (sect. 3). 
Various electromagnetic observables of the
$\Delta -N$ system are calculated and discussed in sect. 4. 
The main results of this work
are summarized in sect. 5. 

\par
\par
\bigskip
\section {\bf The Constituent Quark Model}
\par
\noindent
It is nowadays understood that constituent quarks are quasi-particles,
i.e.  bare quarks surrounded by a polarization cloud of
quark-antiquark pairs that are continuously excited from the QCD
vacuum \cite{Shu84}.
Constituent quarks are therefore complicated objects. They
have a mass $m_q\approx M_N/3$ and a finite hadronic size\footnote{In this 
work we use $r_{q}=0.4$ fm (see eq.(\ref{rpiq})).}.
The constituent quark mass generation is
intimately related to the spontaneously broken chiral symmetry of QCD, i.e.
to the fact that the QCD vacuum is not chirally invariant. The concept
of a massive constituent quark incorporates much of the complexity of
QCD in the low-energy domain of hadron physics. However, there are
still some residual interactions between the constituent quarks.
These simulate those dynamical features of QCD that are not yet
included in the free quasi-particle description.  

\medskip
\subsection{ The Hamiltonian}

\par
\noindent
We consider a baryon as a 
nonrelativistic three-quark system, which, in the case of equal quark
masses $m_q$, is described by the Hamiltonian \footnote{For recent reviews
of the CQM see ref.\cite{Rev91}.}
\begin{equation}
\label{Ham}
H=\sum_{i=1}^{3} \Bigl ( m_q+ {{\bf p}^{2}_{i}\over 2m_{q}} \Bigr )
-{{\bf P}^2\over 6m_q} + \sum_{i<j}^{3} V^{conf}({\bf r}_i,{\bf r}_j)
+
\sum_{i<j}^{3} V^{res}({\bf r}_i,{\bf r}_j),
\end{equation}
where ${\bf r}_i$, ${\bf p}_i$ are the spatial and momentum
coordinates of the i-th quark, respectively. The Hamiltonian of
eq.(\ref{Ham}) consists of the standard nonrelativistic kinetic
energy, a confinement potential and residual interactions $V^{res}$
(see fig.1) which model the relevant properties of QCD.
\par
Asymptotic freedom is modelled in the CQM by the one-gluon exchange
interaction, $V^{OGEP}$, of fig.1(a) which was first introduced by De
Rujula, Glashow and Georgi in 1975
\cite{deR75}:
\bea
\label{gluon}
V^{OGEP} ({\bf r}_i,{\bf r}_j) & = & {\alpha_{s}\over 4}
\b{\lambda}_i\cdot\b{\lambda}_j \Biggl \lbrace
{1\over r}- {\pi\over m_q^2} \left ( 1+{2\over 3}
\b{\sigma}_{i}\cdot\b{\sigma}_{j} \right ) \delta({\bf r})
-{1\over 4m_q^2} {1\over r^3}
\left ( 3\b{\sigma}_i\cdot  {\hat {\bf r} } \,
\b{\sigma}_j\cdot{\hat{\bf r}}
- \b{\sigma}_i\cdot\b{\sigma}_j \right ) \nonumber \\ & & -{1\over
2m_q^2} {1\over r^3} \biggl [ 3 \left ( {\bf r}
\times {1\over2} ({\bf p}_i-{\bf p}_j) \right ) \cdot
{1\over2}(\b{\sigma}_i+\b{\sigma}_j) \nonumber \\ & & - \left ( {\bf
r}\times {1\over2} ({\bf p}_i+{\bf p}_j) \right ) \cdot
{1\over2}(\b{\sigma}_i-\b{\sigma}_j) \biggr ]
\Biggr \rbrace ,
\eea
where ${\bf r}={\bf r}_i-{\bf r}_j$; $\b{\sigma}_i$ is the usual Pauli
spin matrix, and $\b{\lambda}_i$ is the color operator of the i-th
quark.  The one-gluon exchange potential has the correct spin-color
structure of QCD at short distances.
\par
Chiral symmetry is probably the most important feature of
QCD in the nonperturbative regime.  Its importance for hadron physics
has been highlighted in recent reviews \cite{Kir93}.  The
spontaneous breaking of chiral symmetry (SBCS) by the physical vacuum is
responsible for the constituent quark mass generation and the
appearence of pseudoscalar Goldstone bosons ($\pi$-mesons) together
with their massive scalar partners ($\sigma$-mesons) which couple to
the constituent quarks.  In the chiral CQM \cite{Man84}, 
this is modelled in lowest order by introducing one-pion and one-sigma 
exchange potentials between constituent quarks (see fig.1(b-c))
\cite{Shi84,Obu90}:
\bea
\label{Pion}
V^{OPEP}({\bf r}_i,{\bf r}_j) & = & {g_{\pi q}^2\over {4 \pi (4m_q^2)}}
{\Lambda^2\over {\Lambda^2-\mu^2}}
\b{\tau}_{i}\cdot \b{\tau}_{j}\,
\b{\sigma}_{i}\cdot\b{\nabla}_{r}\,
\b{\sigma}_{j}\cdot\b{\nabla}_{r}
\left ( {e^{-\mu r}\over r}- {e^{-\Lambda r}\over r} \right ) \\
V^{OSEP}({\bf r}_i,{\bf r}_j) & = & -{g_{\sigma q}^2\over {4 \pi}}
{\Lambda^2\over {\Lambda^2-m_{\sigma}^2}}
\left ( {e^{-m_{\sigma} r}\over r}- {e^{-\Lambda r}\over r} \right )
\label{VOSEP}
\eea
where $r=\vert {\bf r} \vert = \vert {\bf r}_i-{\bf r}_j \vert $ and
$\mu$ is the pion mass.  Here, $\b{\tau}_i$ denotes the isospin of the
i-th quark.  The parameters of the $\sigma$-exchange potential
$V^{OSEP}$ are fixed by the ones of the $\pi$-exchange potential and
the constituent quark mass \cite{Obu90} via the chiral symmetry constraints
\bea
\label{chiral}
{g^2_{\sigma q}\over {4 \pi}} &= & {g^2_{\pi q}\over {4 \pi}}=
{f^2_{\pi q}\over{4 \pi}} \left ( {2 m_q \over \mu} \right )^2;
\nonumber \\ m_{\sigma}^2 & \approx & (2 m_q)^2+ \mu^2 \nonumber \\
\Lambda_{\pi}& = & {\Lambda_{\sigma}}= \Lambda \equiv \Lambda_{SBCS}.
\eea
The terms in eq.(\ref{Pion}) and eq.(\ref{VOSEP}) involving 
the chiral cut-off $\Lambda$ result from the use of 
$\pi q$ and $\sigma q$ vertex 
functions in momentum space of the form
\be
\label{piqff}
F_{\pi q}({\bf k}^2) =
\left (\Lambda^2 \over \Lambda^2 + {\bf k}^2 \right )^{1/2},
\ee
where ${\bf k}$ is the three-momentum of the pion.  
Thus, the chiral symmetry breaking scale $\Lambda$ is related to the 
hadronic size of the constituent quark via the usual definition
\be
\label{rpiq}
r^2_{\pi q}=-6{d \over d{\bf k}^2} F_{\pi q}({\bf k}^2)
\vert_{{\bf k}^2\equiv 0}
={3 \over \Lambda^2}.
\ee
This finite hadronic size of the constituent quarks is denoted by the
extended vertices in fig.1.  The larger $\Lambda$, the more point-like
the constituent quark.  For $\Lambda \to \infty$ the one-pion exchange
potential of eq.(\ref{Pion}) is unregularized and we recover a
$\delta$-function interaction between point-like constituent quarks.
We usually take for the hadronic size of the constituent quarks
$r_{q}=0.4$ fm, i.e. $\Lambda=4.2$ fm$^{-1}$ in connection with the 
vertex function of eq.(\ref{piqff}). 
\par
It should be emphasized that we introduce $\pi$- and $\sigma$-mesons as
fundamental fields (Goldstone bosons and their chiral partners) 
and not as $q\bar q$ composites. One may also argue that the vertex 
function of eq.(\ref{piqff}) describes {\it both} the hadronic size of 
the pion and of the constituent quarks. 
\par
Finally, in the CQM the confinement of quarks and gluons is
modelled by a linear or quadratic two-body quark-quark potential.
According to Shuryak \cite{Shu84}, the confinement scale is related to
the chiral symmetry breaking scale by
$\Lambda_{conf} \approx \Lambda_{SBCS}$/3. This means that the
distances where confinement effects become important are somewhat
larger than the distances where chiral symmetry is broken. However,
the boundary between the two mechanisms is not very well defined.
Here, we employ a two-body harmonic oscillator confinement potential
\begin{equation}
\label{conf}
V^{conf}({\bf r}_i,{\bf r}_j)= -a_c \b{\lambda}_i\cdot \b{\lambda}_j
({\bf r }_i-{\bf r }_j)^2.
\end{equation}

\bigskip
\goodbreak

\noindent
\subsection{Baryon wave function and determination of parameters} 
\nobreak

\noindent
The total baryon wave function $\Phi_{N(\Delta)}$ is an inner product
of the orbital, spin-isospin, and color wave function and given by
\be
\label{ostcwf}
\mid \Phi_{N (\Delta)}>= 
({1/\sqrt{3} \pi b^2})^{3/2}
\exp(-(\b{\rho}^2/4b^2+\b{\lambda}^2/3b^2))
\mid ST>^{N(\Delta)} \times
\mid [111]>^{N(\Delta)}_{color},
\ee
where the Jacobi coordinates $\b{\rho}$ and $\b{\lambda}$ are defined
as $\b{\rho}={\bf r}_1-{\bf r}_2$ and $\b{\lambda}={\bf r}_3-({\bf
r}_1+{\bf r}_2)/2$.
With the Hamiltonian of eq.(\ref{Ham}) and the wave function of 
eq.(\ref{ostcwf}) it is straightforward to calculate the nucleon mass.
One obtains
\bea
\label{massN}
M_{N}(b)& = & 3m_q+{3\over 2m_q b^2} + V^{conf}(b)
-2\alpha_s \sqrt{{2\over \pi}}{1\over b}
+{1\over 3} { \alpha_s \over m_q^2} {1\over \sqrt{2 \pi}} {1\over b^3} \nonumber \\
& & -{5\over 4} \delta_{\pi}(b)+V^{\sigma}(b), \\
M_{\Delta}(b)&=& 3m_q+{3\over 2m_q b^2} + V^{conf}(b)
-2\alpha_s \sqrt{{2\over \pi}}{1\over b}
+{5\over 3} { \alpha_s \over m_q^2} {1\over \sqrt{2 \pi}} {1\over b^3} \nonumber \\
& & -{1\over 4} \delta_{\pi}(b)+V^{\sigma}(b),
\label{massD}
\eea
where the individual terms in eqs.(\ref{massN},\ref{massD}) are the
nonrelativistic  kinetic energy, 
quadratic confinement, gluon, 
pion, and sigma  contributions, respectively.
The confinement contribution to the nucleon and $\Delta$ mass is given by
\be
V^{conf}(b)=24 a_c b^2, 
\ee
and the $\sigma$-meson potential contribution is 
\be
\label{Vsigma}
V^{\sigma}(b)=-6  {\Lambda^2 \over {\Lambda^2 -m_{\sigma}^2} }
{g^2_{\sigma q}\over 4 \pi}
{1 \over  \sqrt{2\pi}}{1\over b}
\Bigl \{ \Bigl ( 1-\sqrt{\pi}({m_{\sigma} b\over \sqrt{2}})
e^{m_{\sigma}^2b^2/2}
{\rm erfc}({m_{\sigma} b \over \sqrt{2}}) \Bigr )-
(m_{\sigma} \leftrightarrow \Lambda) \Bigr \}.
\ee
Explicit expressions for $\delta_{g}$, $\delta_{\pi}$ are given below.
Subtracting eq.(\ref{massN}) from eq.(\ref{massD}) 
all spin-independent terms drop out and one gets
\be
\label{M-M}
M_{\Delta}-M_N= \delta_g(b)+ \delta_{\pi}(b), 
\ee
where
$\delta_{\pi}(b)$ and $\delta_{g}(b)$ are the {\it spin-dependent} 
pion and gluon contributions to the $\Delta -N$ mass splitting.
%
%
\par
The parameters of the model are: (i) the harmonic oscillator parameter
$b$, (ii) the confinement strength $a_c$, (iii) the strong coupling
constant $\alpha_s$, (iv) the cut-off mass $\Lambda$ in the pion-quark
and sigma-quark interaction.  
For the constituent quark mass we choose $m_q=M_N/3=313$ MeV.  The parameters
$a_c$, $\alpha_s$, and $b$ are determined from the three conditions
\be
\label{constr}
M_{N}(b)=3m_q= 939 {\rm MeV}, \qquad M_{\Delta}-M_{N}=
\delta_{\pi}(b)+\delta_g(b)=293 {\rm MeV}, \qquad {\partial M_{N}(b)
\over \partial b}= 0,
\ee
as previously described \cite{Buc91}. The gluon ($\delta_g$) and 
pion ($\delta_{\pi}$) contributions to the $\Delta -N$ mass splitting
are calculated as
\be
\label{deltag}
\delta_g (b)= {4 \alpha_s  \over 3 \sqrt{2\pi} m_q^2  b^3 },
\ee
\be
\label{deltapi}
\delta_{\pi}(b)=
-4 {\Lambda^2 \over {\Lambda^2 -\mu^2} }
{f^2_{\pi q} \over {4 \pi} \mu^2} \sqrt{2\over \pi}{1\over b}
\Bigl \{ \mu^2 \Bigl ( 1-\sqrt{\pi}({\mu b\over \sqrt{2}})e^{\mu^2b^2/2}
{\rm erfc}({\mu b \over \sqrt{2}}) \Bigr )-(\mu \leftrightarrow \Lambda) \Bigr \},
\ee
respectively. In the following we write
$\delta_{\pi}(b)=\delta_{\pi_{\mu}}(b)-\delta_{\pi_{\Lambda}}(b)$ 
for brevity.
Numerical values for the individual contributions to the nucleon mass
and to the $\Delta-N$ mass splitting  are listed in table 2.

\par
The residual interactions will  
admix  higher excited states to the pure $(0s)^3$ ground state
wave functions of eq.(\ref{ostcwf}) (configuration mixing).
If we restrict ourselves to $2\hbar \omega$ excitations,
we have four excited states
($\Phi^N_{S'_{S}}$, $\Phi^N_{S_{M}}$, $\Phi^N_{D_{M}}$, 
$\Phi^N_{P_{A}}$) for the $N$
and three excited states
($\Phi^{\Delta}_{S'_{S}}$, $\Phi^{\Delta}_{D_{S}}$, $\Phi^{\Delta}_{D_{M}}$)
for the $\Delta$.
The subscripts $L_{sym}$ describe the orbital angular momentum
($L$) and the symmetry ($sym$) of the orbital wave function under particle
exchange. Here, $S$ denotes symmetric, $M$ mixed symmetric 
and $A$ antisymmetric orbital wave functions.  
The $N$ and $\Delta$ wave functions are then given by
\begin{eqnarray}
\label{cfm}
\Phi_N & = &  a_{S_{S}} \Phi^N_{S_{S}} 
+a_{S'_{S}} \Phi^N_{S'_{S}} +a_{S_{M}} \Phi^N_{S_{M}}
+a_{D_{M}} \Phi^N_{D_{M}} +a_{P_{A}} \Phi^N_{P_{A}} 
\nonumber \\
\Phi_{\Delta} & = &  b_{S_{S}} \Phi^{\Delta}_{S_{S}} 
+b_{S'_{S}} \Phi^{\Delta}_{S'_{S}} 
+b_{D_{S}} \Phi^{\Delta}_{D_{S}}
+b_{D_{M}}  \Phi^{\Delta}_{D_{M}}. 
\end{eqnarray} 
A detailed description of these wave functions 
can be found in ref.\cite{Gia90}. 

\par
In ref.\cite{Buc91,Buc94} we used the wave functions of eq.(\ref{cfm}) 
and simultaneously calculated the positive parity excitation spectrum of 
the $N$ and $\Delta$ and the electromagnetic properties of the nucleon.
In most cases this slightly improved the results for the electromagnetic 
properties in comparison to a pure ground state calculation. 
The main effect of configuration mixing was to increase the 
pion contribution and to reduce the gluon contribution to various observables. 
In addition, the value of the harmonic oscillator 
constant $b$ was slightly reduced with respect to a pure $L=0$ ground state 
calculation.

\par
In this work, we use the ground state wave 
function of eq.(\ref{ostcwf}) since this considerably simplifies calculations.
Because the mixing amplitudes, as obtained by different groups 
\cite{Kon80,Dre84,Clo90,War90,Buc91} are small, 
our calculation provides the dominant part of the 
exchange current contribution to different observables. 
Certainly, the simple relations between different low energy observables, which
we will derive in sect. 4 do not hold exactly in a more
complete calculation with configuration mixed wave functions.
Still, we expect them to hold in good approximation,
for example,  $r_n^2=-b^2(M_{\Delta}-M_N)/M_N$ holds true at the level 
of $23\%$ or better even if configuration mixed wave functions are used
\cite{Buc91}. 
Therefore, we think that our conclusions concerning the role
of exchange currents will remain true also in a more consistent calculation
employing the full wave functions of eq.(\ref{cfm}). 
We will discuss this in more  detail in sect. 4.

%
\bigskip
\goodbreak
\section {Electromagnetic currents }

\nobreak
\par
\noindent
The interaction of the external electromagnetic field
$A^{\mu}(x)=(\Phi(x),{\bf A}(x))$ with a hadronic system is described
by the Hamilton operator
\be
\label{hem}
H_{em}=\int \, d^4 x \ J_{\mu}(x) A^{\mu}(x).
\ee  
where $J_{\mu}(x)=(\rho(x),-{\bf J}(x))$ is the four-vector current
density of the quarks inside the system.  Thus, in a quark model
description we must know the charge and current operators of the
interacting quarks in order to describe the electromagnetic properties
of the baryon.

\goodbreak
\subsection{ One-body current}
\nobreak
\par
\noindent 
First, we consider the standard nonrelativistic one-body charge and
current operators of point-like constituent quarks (see fig.2a)
\begin{eqnarray}
\label{imp}
{\rho}_{imp}^{IS/IV} ({\bf r}_i,{\bf q)} & = & {e_i} e^{i{\bf{q \cdot
r }}_i} \nonumber \\
{\bf J}_{imp}^{IS/IV} ({\bf r}_i,{\bf q)} & = & {e_i\over2m_q} \left(
i \left \lbrack \b{\sigma}_i\times {\bf p}_i,e^{i{\bf{q \cdot r }}_i}
\right \rbrack +
\left \lbrace {\bf p}_i, e^{i{\bf{q \cdot r }}_i} \right \rbrace \right ),
\end{eqnarray} 
where $e_i={1\over 6}e(1+3{\b{\tau_i}}_3)$ is the quark charge
operator and ${\bf q}$ is the 
three-momentum transfer of the photon.
  Here and in the following $ {\b{\tau_i}}_3$ denotes the
third component of the isospin of the $i$-th quark.  The
decomposition of eq.(\ref{imp}) into isoscalar (IS) or isovector (IV)
currents is obtained by taking only the first or second term of the
quark charge operator into account. 
Note that we do not use any anomalous magnetic moments for the constituent
quarks \cite{Wei90}.
\goodbreak

\subsection{ Two-body exchange currents} 
\par
\nobreak
\noindent
In most applications of the CQM the total electromagnetic current has
been approximated by a sum of single-quark currents of the form of
eq.(\ref{imp})
\be
\label{impapp}
J_{total}^{\mu}\approx \sum_{i=1}^3 J_{imp}^{\mu}(i).
\ee
\noindent
However, the current of eq.(\ref{impapp}) is not conserved in the
presence of various residual interactions between the quarks.  In a
bound system of quarks the electromagnetic current operator is not
simply a sum of free quark currents as in eq.(\ref{impapp}) but
necessarily contains various two-body currents for the total
electromagnetic current to be conserved.  The spatial parts of these
two-body currents are closely related to the quark-quark potentials
from which they can be derived by minimal substitution \cite{Buc94}.
\par
In the following, we list the two-body
charge and current operators employed in this work.  They have been
derived by a nonrelativistic expansion of the Feynman diagrams of
fig.2(b-e) up to lowest nonvanishing order. 
Only, in the case of the isovector pion pair-current 
we also list the next-to-leading order term
for reasons discussed in sect. 4.3.
For the gluon and pion
exchange currents we obtain \cite{Buc91,Buc94}
\begin{eqnarray}
\label{gpair}
%
%
%
%
%
{ \rho}_{gq{\bar q}}^{IS/IV} ({\bf r}_i,{\bf r}_j,{\bf q})& = &
-i{{\alpha_s}\over{16 m_q^3 }}\,{\b{\lambda}_i}\cdot{\b{\lambda}_j}
\left \lbrace  e_ie^{i{\bf q}\cdot {\bf r}_i} 
\left [ {\bf q} \cdot {\bf r} + 
({\b {\sigma}_i}\times{\bf q}) \cdot ({\b {\sigma}_j}\times{\bf r })
\right ] +(i\leftrightarrow j) 
\right\rbrace {1\over r^3},  \nonumber \\
%
%
%
%
{\bf J}_{gq{\bar q}}^{IS/IV} ({\bf r}_i,{\bf r}_j,{\bf q})& = &
-{{\alpha_s}\over{4 m_q^2 }}\,{\b{\lambda}_i}\cdot{\b{\lambda}_j}
\Bigl\lbrace {e_ie^{i{\bf q}\cdot {\bf r}_i}}{1\over 2}
({\b {\sigma}_i}+{\b {\sigma}_j})\times{\bf r} +(i\leftrightarrow
j)\Bigr\rbrace {1\over r^3};
\end{eqnarray} 

\bea
%
%
%
%
%
\label{ppis}
{\rho}^{IS}_{\pi q {\bar q}} ({\bf r}_i,{\bf r}_j,{\bf q}) & = &
{ie\over 6} {g^2_{\pi q}\over 4\pi (4m_q^3)} {\Lambda^2 \over
\Lambda^2-\mu^2}
\,\b{\tau}_i \cdot \b{\tau}_j
\left \{ 
e^{i{\bf q}\cdot {\bf r}_i} \b{\sigma}_i \cdot {\bf q} \,
\b{\sigma}_j\cdot{\bf {\nabla_r}} +(i\leftrightarrow j) 
\right \}
\left ( {e^{-\mu r}\over r}- {e^{-\Lambda r}\over r}  \right ), \nonumber \\
%
%
%
{\bf J}^{IS}_{\pi q {\bar q}}({\bf r}_i,{\bf r}_j,{\bf q}) & = &
{ie\over 6}{g^2_{\pi q}\over 4\pi (8m_q^4) } {\Lambda^2 \over
\Lambda^2-\mu^2}
\, \b{\tau}_i\cdot \b{\tau}_j
\left \{ 
e^{i{\bf q}\cdot {\bf r}_i} {\bf q}\times{\bf \nabla_r}\,
\b{\sigma}_j \cdot {\bf {\nabla_r}}+(i\leftrightarrow j)
\right \} 
\left ( {e^{-\mu r}\over r}- {e^{-\Lambda r}\over r} \right ).  \nonumber \\
& &
\eea
\bea
%
%
%
%
%
\label{ppiv}
%
%
{\rho}^{IV}_{\pi q {\bar q}}({\bf r}_i,{\bf r}_j,{\bf q}) & = &
{ie\over 2} {g^2_{\pi q}\over 4\pi (4m_q^3) } {\Lambda^2 \over
\Lambda^2-\mu^2} \,
\left \{ 
{\b{\tau}_j}_3 e^{i{\bf q}\cdot {\bf r}_i} \b{\sigma}_i \cdot {\bf q}
\,
\b{\sigma}_j \cdot {\bf {\nabla_r}}+(i\leftrightarrow j) 
\right \}
\left ({e^{-\mu r}\over r}- {e^{-\Lambda r}\over r} \right ), \nonumber  \\
%
%
{\bf J}^{IV}_{\pi }({\bf r}_i,{\bf r}_j,{\bf q})& = & e{g^2_{\pi
q}\over 4\pi (2m_q)^2} {\Lambda^2 \over \Lambda^2-\mu^2} 
\Biggl [
\left \{
\,({\b{\tau}_i}\times {\b{\tau}_j})_3
{e^{i{\bf q}\cdot {\bf r}_i}} \b{\sigma}_i \
\b{\sigma}_j \cdot{\bf {\nabla_r}}+(i\leftrightarrow j)
\right \}
\left ( {e^{-\mu r}\over r}- {e^{-\Lambda r}\over r} \right )    \nonumber \\
& & +
{i\over 4 m_q^2} \, 
\left \{    {{\b{\tau}}_j}_3
e^{i{\bf q}\cdot {\bf r}_i} {\bf q}\times{\bf \nabla_r}\,
\b{\sigma}_j \cdot {\bf {\nabla_r}}+(i\leftrightarrow j)
\right \} 
\left ( {e^{-\mu r}\over r}- {e^{-\Lambda r}\over r} \right ) \nonumber \\  
& & + \,({\b{\tau}_i}\times {\b{\tau}_j})_3
\b{\sigma}_i \cdot  {\bf \nabla}_i \, 
\b{\sigma}_j \cdot  {\bf \nabla}_j
\int_{-1/2}^{1/2}\!\!\!\!\! dv 
e^{i{\bf q}\cdot ({\bf R}-{\bf r}v)}
\left ( {\bf z}_{\mu} { e^{-L_{\mu }r}\over L_{\mu} r}-
{\bf z}_{\Lambda} {e^{-L_{\Lambda }r}\over L_{\Lambda} r} \right ) 
\Biggr ]. 
\eea
The first two terms in eq.(\ref{ppiv}) are the leading order pion 
pair-current proportional to $({\b{\tau}_i}\times {\b{\tau}_j})_3$
and its next-to-leading order relativistic correction 
proportional to ${\b{\tau}_{j}}_3$ shown in fig. 2b. 
The third term in eq.(\ref{ppiv}) is the pionic current
${\bf J}^{IV}_{\gamma \pi \pi } $ of fig. 2c. We have used the following
abbreviations 
 ${\bf R}=
({\bf r}_i+{\bf r}_j)/2 $, ${\bf z}_m({\bf q},{\bf r})=L_m{\bf
r}+ivr{\bf q}$, and $L_{m}(q,v) = [{1\over 4}q^2(1-4v^2)+m^2]^{1/2} $.
\par
The scalar pair-current corresponding to a Lorentz-scalar interaction
was previously \cite{Buc94} derived as:
\begin{eqnarray}
\label{scalar}
%
%
%
%
\rho^{IS/IV}_{scalar}({\bf r}_i,{\bf r}_j,{\bf q}) & = &
{1\over (2m_q)^3}
\left \{ 
e^{i{\bf q}\cdot {\bf r}_i}e_i
\left ( {3\over2} {\bf q}^2-i{\bf q}\cdot \nabla_r
+{1\over 2}\nabla_r^2 \right ) V^{scalar}({\bf r}_i,{\bf r}_j)
+(i\leftrightarrow j)
\right \}
\nonumber \\
%
%
{\bf J}^{IS/IV}_{scalar }({\bf r}_i,{\bf r}_j,{\bf q}) & = & -{1\over
2m_q^2}
\Bigl \{  e_i e^{i {\bf q}\cdot {\bf r}_i}
\b{\sigma}_i \times {\bf q}   V^{scalar}({\bf r}_i,{\bf r}_j) 
+(i\leftrightarrow j) \Bigr \}.
\end{eqnarray} 
Eq.(\ref{scalar}) is used to calculate both the confinement- and
$\sigma$-meson-exchange currents. To obtain the spatial part
of this current directly from the potential one must
reduce the relativistic scalar potential to the same order 
in $1/m_q^2$ as the one-gluon exchange potential. 
By minimal substitution ${\bf p}_i\to {\bf p}_i- e{\bf A}({\bf r}_i) $
in the scalar potential and by adding the contribution of the commutator 
of the $ {\cal O}(1/m_q^2)$ term in the one-body charge density with the 
leading order scalar potential one obtains the scalar pair-current shown 
in fig.2e. This is explained in greater detail in ref. \cite{Buc94}. 
\par
The total charge operator consists of the usual one-body charge operator
and two-body charge operators due to the interaction between
the quarks
\be
\label{totcha}
\rho^{}_{total}({\bf q})= \sum_{i=1}^3 \rho^{}_{imp}({\bf r}_i) +\sum_{i<j}^3
 \left (       \rho^{}_{g q\bar q}({\bf r}_i, {\bf r}_j)
               +\rho^{}_{\pi q\bar q}({\bf r}_i, {\bf r}_j)
               +\rho^{}_{\sigma q\bar q}({\bf r}_i, {\bf r}_j)
               +\rho^{}_{conf}({\bf r}_i, {\bf r}_j) \right ) .
\ee
Likewise the total current operator consists of the usual one-body  
operator and two-body exchange current operators 
tightly related to the different quark-quark interactions
\be
\label{totcur}
{\bf J}^{}_{total}({\bf q}) = \sum_{i=1}^3 {\bf J}^{}_{imp}({\bf r}_i) +
\sum_{i<j}^3 \left (  {\bf J}^{}_{g q\bar q}( {\bf r}_i, {\bf r}_j) 
 +{\bf J}^{IV}_{\pi}( {\bf r}_i, {\bf r}_j) 
 +{\bf J}^{IS}_{\pi q\bar q}( {\bf r}_i, {\bf r}_j) 
 +{\bf J}^{}_{\sigma q\bar q} ( {\bf r}_i, {\bf r}_j) 
 +{\bf J}^{}_{conf}( {\bf r}_i, {\bf r}_j) \right ). 
\ee
The extent to which the spatial current
satisfies the continuity equation with the potential 
used in sect. 2 has been discussed previously \cite{Buc94}.

\par 
\goodbreak

\subsection{Electromagnetic size of the constituent quarks}
\nobreak 
\noindent
In sect. 2 we have seen that constituent quarks
have a finite hadronic size which is given by the 
hadronic form factor of eq.(\ref{piqff}). 
Similarly, the {\it electromagnetic} size of the constituent quarks is 
described by a monopole form factor 
\be
\label{qemff}
F_{\gamma q}({\bf q}^2) =
 {1 \over 1  + {1\over 6} {\bf q}^2 r^2_{\gamma q} }. 
\ee
In order to take the internal electromagnetic structure of the 
constituent quarks into account,
the charge and current operators of the previous section 
must be multiplied by the form factor of eq.(\ref{qemff}).

\par 
The finite electromagnetic radius $r_{\gamma q}$ takes
into account that constituent quarks are dressed particles, i.e.
current quarks surrounded by a cloud of $q \bar q$-pairs. 
The dominant contributions come from 
quark-antiquark pairs with pion quantum numbers. Vector meson 
dominance relates the electromagnetic radius of the constituent quarks 
to the $\rho$-meson pole according to fig.3. 
The notion of a finite electromagnetic size of the 
constituent quarks has been used before \cite{Pet81,Sta82,Vog90}.
While the mass and size of the constituent quarks are appreciably
renormalized from the point particle values 
explicit calculation in the Nambu-Jona-Lasinio model 
shows that the anomalous magnetic moment of constituent quarks
is small \cite{Vog90}. This was previously anticipated on 
general grounds \cite{Wei90}.


\bigskip

\goodbreak 

\bigskip
\noindent
\goodbreak
\section {\bf $\Delta$ electromagnetic properties}
\nobreak
\smallskip
\nobreak
\noindent
A hadron with spin $J$ has, in general, $2J+1$ elastic electromagnetic form
factors. This result can be deduced by writing the most 
general Lorentz-invariant expression for the electromagnetic current
operator of a hadron with total angular momentum $J$.
One then demands hermiticity, and that the diagonal matrix
elements be invariant under time and parity transformations and
satisfy the continuity equation for the electromagnetic current. 
This reduces the number of allowed form factors to $2J+1$.  The $\Delta$
thus has four elastic form factors \cite{Are76} (and references therein): 
the charge monopole $F_{C0}$, the charge  quadrupole $F_{C2}$,
the magnetic dipole $F_{M1}$, and the magnetic octupole $F_{M3}$.
It turns out that the $M3$ form factor vanishes exactly 
for the ground state wave functions considered in
this work. In order to describe the electromagnetic $N \to \Delta$ transition
\cite{Jon73} one needs the transverse magnetic 
dipole $F_{M1}^{ N \to \Delta }$, 
the transverse electric quadrupole 
$F_{E2}^{N \to \Delta }$
 and the charge quadrupole (longitudinal) 
$F_{C2}^{ N \to \Delta }$ transition form factors. 

\par
In this work we concentrate on electromagnetic moments and radii
of the $\Delta$, where the nonrelativistic quark model is expected 
to work best. Unlike the full form factors, the results for 
these static properties can be obtained in terms of analytic 
expressions, which makes the relation between the 
electromagnetic properties of the nucleon and $\Delta$ more transparent. 
In particular, the important role of non-valence quark degrees of freedom 
in various electromagnetic properties will become evident. 
A review of $\Delta$ electromagnetic properties 
in the quark model has been given by Giannini \cite{Gia90}.

\par 
\goodbreak
\subsection{Charge Radii }
\nobreak 
\noindent

Charge radii measure the spatial extension of the charge 
distribution inside the baryon. They contain information 
about non-valence quark degrees of freedom and about the finite 
electromagnetic size of the valence quarks.  
Quite generally, the charge radius is defined as the slope of the charge 
form factor at zero-momentum transfer 
\be
\label{chr}
r^2_C= -{6 \over F_C(0)} {d \over d { {\bf q}}^2 } F_{C}({\bf q}^2)\mid_{{\bf q}^2=0},
\ee
where, according to the general definition of the elastic 
form factors \cite{Gar76},
\be
\label{chaff}
F_C({\bf q}^2)= {\sqrt{4\pi}} < J M_J\! \!=\! \!J \, \, T M_T \mid 
         {1\over 4\pi} \int d\Omega_q \rho({\bf q}) Y^0_0(\hat{\bf q})
         \mid J M_J\!\!=\!\!J \,\, T M_T>.
\ee

\par 
\goodbreak
\subsubsection{Charge Radius of the $\Delta$}
\nobreak 
\noindent 
\par

Using eq.(\ref{totcha}) and the ground state wave functions of 
eq.(\ref{ostcwf}), we obtain the following analytic expressions 
\be
\label{rdel}
r^2_{\Delta} = b^2 + r_{\gamma q}^2+ {b^2\over 6 m_q} (5\delta_g -
\delta_{\pi}) + {5 \over 6 m_q^3} V^{conf }+ 
r_{\sigma}^2.
\ee
Eq.(\ref{rdel}) is valid for charged $\Delta$ states,  
while $r^2_{\Delta^0}$ is zero in the present model. 
The first two terms in eq.(\ref{rdel}) are due to the one-body quark 
current including the finite electromagnetic size of the quarks, 
while the remaining terms represent the gluon, 
pion, confinement, and sigma exchange current contributions.  
An analytic expression for $r_{\sigma}^2$ 
can be obtained by replacing $V^{conf}$ 
by $V^{\sigma}$ in eq.(4.11) of ref. \cite{Buc94}.
Note that the spin-independent scalar two-body charge densities have the same 
isospin structure as the one-body charge density.  
Therefore, as in the case of the one-body charge density contribution,
the spin-independent scalar exchange current contributions
to the charge radii are identical for $N$ and $\Delta$.
Our numerical results are listed in table 3.
\par
If we compare this with the corresponding result for the proton
\be
\label{rpmec}
r^2_{p}  =  b^2 + r_{\gamma q}^2+ {b^2\over 2 m_q} (\delta_g -
\delta_{\pi}) + {5 \over 6 m_q^3} V^{conf }+ 
r_{\sigma}^2
\ee
and neutron 
\be 
\label{rnmec}
r^2_{n}  = 
-{b^2\over 3m_q} (\delta_g+ \delta_{\pi}) = -b^2 {M_{\Delta}- M_N
\over M_N},
\ee
we obtain 
from eqs.(\ref{rdel},\ref{rpmec}, \ref{rnmec}) 
the parameter-independent result
\be 
r^2_{\Delta}=r^2_p-r^2_n.
\ee
Hence, the charge radius of the $\Delta$ is equal to the isovector 
charge radius of the nucleon. Stated differently, 
the charge radius of the $\Delta$ is somewhat larger than that
of the proton, and the difference is given by the neutron charge
radius. This is in agreement with other models of nucleon structure
\cite{Gob92}. 
\par
\par
We have noted that the charge radius 
of the $\Delta^0$ is exactly zero in this model. 
This is so, because all terms in eq.(\ref{totcha}) 
yield contributions to the $\Delta$  charge form factor 
which are proportional to the $\Delta$ charge 
\be
\label{chadel} 
e_{\Delta}={1\over2} (1 + 2 M_T), 
\ee
where $M_T$ is the third component of the isospin of the $\Delta$.
Therefore, the form factor of eq.(\ref{chaff}) vanishes identically and 
the corresponding charge radius is zero.
\par
In contrast to this, the neutron charge radius of eq.(\ref{rnmec})
is clearly nonzero, but in this case we also obtain a particularly 
simple result. Because the neutron charge radius is given
by the difference of isoscalar and isovector radii, the contributions
of the one-body charge density, the finite size of the quarks, and  
the spin-independent scalar (confinement and sigma) exchange currents all 
cancel in $r^2_n$.  
Only the spin-dependent pion and gluon exchange currents contribute 
to $r^2_n$. The gluon and pion exchange currents can be expressed in terms of 
$\delta_{\pi}$ and $\delta_g$, i.e. the pion and gluon contribution to 
the $N-\Delta$ mass splitting, because the exchange current operators have a 
structure similar to the corresponding potentials. 
The particular combination
of $\delta_{\pi}$ and  $\delta_g$ appearing in eq.(\ref{rnmec}) 
makes it possible to express $r^2_n$ via the experimental $\Delta -N$
mass splitting of eq.(\ref{constr}).  
Eq.(\ref{rnmec}) clearly shows
that there is an intimate relation between: (i) the neutron charge
radius, (ii) the spatial extension of the quark distribution inside
the nucleon (the quark core radius $b$), and (iii) the excitation
energy to the first excited state of the nucleon.  From
eq.(\ref{rnmec}) we determine the quark core size as $b=0.612$ fm, if
the experimental numbers for $M_N$, $M_{\Delta}$, and $r^2_n$ are
substituted.

\par 
\goodbreak
\subsubsection{Configuration mixing vs. exchange currents}
\nobreak 
\noindent 
\par

Let us try to give a physical interpretation of the results 
of eq.(\ref{rdel},\ref{rnmec}). In previous works, 
the nonvanishing charge radius of the 
neutron was attributed to the perturbing effect of the color-magnetic 
interaction on the ground state wave function \cite{Car77,Isg82}. 
The color-magnetic interaction provides a repulsive force between any two 
quarks which are in a symmetric spin state ($S=1$).  This makes the 
$\Delta$-isobar heavier than the nucleon, since the former contains more spin
symmetric quark pairs.  Similarly, the color-magnetic force repels the
two down quarks in the neutron which are necessarily in an $S=1$ state
(Pauli principle). This leads to a negative tail in the neutron charge
distribution and to a negative neutron charge radius. 
On the other hand, the $\Delta^0$ is symmetric in spin space 
and the spin-dependent forces do not introduce any asymmetry between 
$ud$ and $dd$ quark pairs. Therefore, the charge radius of the 
$\Delta^0$ is zero.
Thus, the same physical mechanisms (one gluon- and one-pion exchange) are 
responsible for the $\Delta-N$ mass splitting and the negative charge radius 
of the neutron.

We point out that this effect, which is usually described by a small admixture 
of the excited $\Phi_{S_M}$ state of eq.(\ref{cfm})
into the nucleon ground state wave function, {\it is
much too small}. It is around $r_n^2(imp)=-0.03$ fm$^2$ 
if a realistic quark core radius ($b \approx 0.6$ fm )
is used (see the discussion below fig.2 in ref. \cite{Buc91}). 
The success of previous impulse calculations for the neutron charge 
is bought by tolerating a severe inconsistency: a value of 
$b \approx 0.5-0.6$ fm is typically used in the calculation of the excitation 
spectrum \cite{Ger82} while a value $b \approx 1 $ fm is employed in the 
neutron charge radius calculation \cite{Gia90}.  In addition, a large quark 
core radius $b \approx 1$ fm contradicts information from several other 
sources and seems to be ruled out \cite{Gia90}. 

The present explanation of the negative neutron
charge radius is based on the spin-dependent two-body gluon- and pion
{\it exchange current operators}. This allows to get the correct size of the
neutron charge radius for a reasonably small quark core radius 
$b \approx 0.6$ fm.
The exchange currents which we discuss here are closely related to 
the spin-dependent terms in the potential which give rise 
to the $\Phi_{S_M}$ state. Yet, there is an important difference between these 
two mechanisms. We will explain this in more detail in the next section.

Of course, in a fully consistent calculation both configuration mixing
and exchange currents must be included 
and the question concerning their relative importance arises.
In addition, the simple relation between the neutron charge radius
and the $N-\Delta$ mass splitting will be modified in a more consistent
calculation. However, according to ref.\cite{Buc91} eq.(\ref{rnmec}) is 
satisfied to within $23\%$ in a model with gluons only; in the model with
gluons and pions it holds to within $12\%$ even if configuration
mixing is included. Therefore, we believe that eq.(\ref{rnmec}) correctly 
describes the physics behind the nonvanishing neutron charge radius
and certain other observables, that are very sensitive to nonvalence
quark degrees of freedom (gluons, pions, and sea-quarks). 

In our previous calculation of the neutron charge form factor \cite{Buc91}
including both configuration mixing and exchange currents we have seen
that the neutron charge radius is clearly dominated by the 
gluon and pion quark-pair currents (see fig.2b-c) 
if a reasonably small quark core radius ($b=0.5-0.6$ fm) is used. 
This finding gets support from other
sources. For example, Christov {\it et} al. \cite{Chr96} find in their chiral 
Nambu-Jona-Lasinio type quark model that the neutron charge radius is 
completely dominated  by sea-quarks and not by valence quark degrees of 
freedom. 
In the language of quark potential models,
it is most natural to include these nonvalence quark degrees of freedom in
electromagnetic observables in the form of {\it gluon and pion exchange
currents}.

\par
\par
\goodbreak 
\subsection{Quadrupole moments }
\nobreak

\noindent
QCD predicts effective tensor forces between quarks. Consequently,
baryons should be deformed.  Experiments at all major electron
laboratories are being devoted to measuring this deformation by
photo/electro-excitation of the $\Delta$-resonance \cite{Eri94}.
From these measurements one
hopes to extract the $D$-state probabilities $a_D^2$ and $b_D^2$ 
in eq.(\ref{cfm}) and from these further information 
about the tensor force between quarks.  

\par
On the other hand, it is well known from nuclear physics that rigorous
bounds on D-state admixtures are difficult to obtain from observables
such as quadrupole and magnetic moments.  For example, in the case of
the deuteron, meson exchange current corrections destroy the direct
relation between the D-state admixture and the measured magnetic
and quadrupole moments. Before we can extract information about interquark 
forces from these observables, we must have some knowledge of the effect of
exchange currents. 

\par
The quadrupole moment is defined as the $q\to0$ limit of the quadrupole
form factor \cite{Gar76}
\be
\label{quadff}
F_Q({\bf q}^2)= -{12 \sqrt{5\pi} \over q^2} 
< J M_J\!\!=\!\!J \,\, T M_T \mid 
         {1\over 4\pi} \int d\Omega_q \, \rho({\bf q}) \, Y^2_0(\hat{\bf q})
         \mid J M_J\!\!=\!\!J \,\, T M_T>,
\ee 
where $J=T=3/2$.

\par
\par
\goodbreak 
\subsubsection{Quadrupole moment of the $\Delta$ }
\nobreak

With the two-body charge densities
employed in this work we derive a parameter-independent relation
between the neutron charge radius and the quadrupole moment of the
$\Delta$
\be
\label{quad}
Q_{\Delta}=-b^2{(\delta_g+\delta_{\pi})\over 3 m_q} e_{\Delta} 
= 
-b^2 \left ( {M_{\Delta}-M_N \over M_N} \right ) e_{\Delta} =r^2_n
e_{\Delta},
\ee
where $e_{\Delta}=(1+2M_T)/2$ is the charge of the $\Delta$.  
Hence, for the $\Delta^{++}$ we predict a quadrupole moment of 
$Q_{\Delta^{++}}=-0.235$ fm$^2$. Numerical values for the other 
quadrupole moments are listed 
in table 4. A similar relation between the neutron charge radius and the 
$\Delta$ quadrupole moment has been obtained on the basis of configuration 
mixing and a $Q_{\Delta^{++}}=-0.093$ fm$^2$ has been 
found \cite{Kri91}. We will discuss the relation between 
exchange current and configuration mixing (tensor force) 
contributions to $Q_{\Delta}$ in more detail below.

\par  
The first thing one notices is that even without an explicit $D$-state 
admixture
in the $\Delta$ wave function, we have obtained a nonvanishing 
quadrupole moment.
In the following, we provide an explanation for this result.
According to the definition of the quadrupole form factor
of eq.(\ref{quadff}), the charge 
density operator must contain a term
$Y^{[2]}(\hat{\bf q})$, 
otherwise the quadrupole form factor vanishes, due to the orthogonality
of the spherical harmonics. For example, after expanding 
the plane wave in eq.(\ref{imp}), 
the one-body charge operator is proportional to
\be
\rho^{imp}_{[1]} \propto
\left [ Y^{[l]}(\hat{\b{\rho}}) \times Y^{[l]}(\hat{\bf q}) \right ]^{[0]}. 
\ee
For a pure $S$-state $\Delta$ wave function only the term $l=0$ 
can contribute and consequently a $Y^{[2]}(\hat{\bf q})$ term is not allowed.
On the other hand, the two-body gluon and pion 
charge densities contain a rank 2 tensor in spin space
\be
\label{rank2}
\rho^{exc}_{[2]} \propto
  \left [ [ \b{\sigma}^{[1]}_i \times  \b{\sigma}^{[1]}_j ]^{[2]}
    \times [ Y^{[l]}(\hat{\b{\rho}}) \times Y^{[2]}(\hat{\bf q}) ]^{[2]} 
\right ]^{[0]}.
\ee
Therefore, it is possible to have a $Y^{[2]}(\hat{\bf q})$ part
even if $l=0$ and the quarks are all in S-states. 
That is why the two-body charge densities derived from fig.2b and fig.2d 
lead to a nonvanishing quadrupole moment.
To state this in more physical terms we can say that 
due to the spin-dependent interaction currents between the quarks, 
the system can absorb a $C2$  or $E2$ photon.

\par  
As is evident from eqs.(\ref{quad}), these two-body charge densities 
describe the same gluon and pion degrees of freedom which are responsible 
for the tensor forces between quarks. 
The physical interpretation of both types of contributions
(tensor force vs. two-body current) to the quadrupole moment is, however, 
quite different.
This is illustrated in fig.4(a-b). The {\it two-body} gluon and pion 
pair-charge densities of fig.4b describe, as their name implies, 
the excitation of quark-antiquark pairs by the photon, or, stated differently,
the absorption of a $C2$ photon on {\it two quarks}. 
On the other hand, in fig.4a the photon is absorbed by a {\it single quark}, 
which remains in a positive energy state between the 
absorption of the photon and the emission of the gluon or pion. 
There is no electromagnetic coupling of the photon
to the quark-antiquark pairs inside the $\Delta$ in this case.
In fig.4a, gluon and pion degrees of freedom show up 
as tensor force induced $D$-state admixtures to ground state 
wave functions (see eq.(\ref{cfm})). 
In most applications of the CQM, the single-quark current of fig.4a 
has been used to estimate the effect of the one-gluon exchange 
potential on electromagnetic properties. Our results show that this
is not a good approximation for the charge properties of the $\Delta$,
which are appreciably affected by exchange currents. 
This is opposite to what one finds in light nuclei.
For example, the deuteron quadrupole moment is mainly
caused by the $D$-wave in the deuteron and  
exchange currents, for example, the pion-pair charge density in eq.(25), 
provide only a correction of about $4\%$ \cite{Buc89}.

Obviously, a complete calculation comprises both 
$D$-waves in the nucleon and the exchange currents discussed in this
work. Corrections due to $D$-waves will modify the simple result
of eq.(\ref{quad}). However, according to our previous experience with 
the neutron charge radius, we expect it to remain largely valid. 
Let us discuss this more quantitatively. 
Including configuration mixing but no exchange currents one obtains 
neglecting the small $b_D^2$ 
contributions and with typical values for the admixture coefficients
\cite{Ger82,Gia90} 

\be
Q_{\Delta}= -b^2 \, {4 \over \sqrt{30} }\left ( b_{S_S} b_{D_S} 
+ {2\over \sqrt{3} }
b_{S'_S} b_{D_S} \right ) e_{\Delta}=-0.087 b^2 e_{\Delta}. 
\ee
For $b=0.61$ fm one obtains then 
$ Q^{imp}_{\Delta}= -0.032 {\rm fm}^2 \, e_{\Delta}$.
This has to be compared to our result 
$ Q^{exc}_{\Delta}= -0.119 {\rm fm}^2 e_{\Delta}$. 
Thus, in a more complete calculation we expect the corrections to
eq.(\ref{quad}) coming from configuration mixing to be below 
some 30 $\%$. In any case, our results clearly indicate that 
an eventual measurement of the quadrupole moment of the $\Delta$ 
should not be interpreted in terms of an 
intrinsic deformation ($D$-waves) alone; 
it is more likely that quark-antiquark pair currents provide 
the dominant contribution to the quadrupole moment of the $\Delta$.

\par
\par
\goodbreak 
\subsubsection{$N \to \Delta$ transition quadrupole moment }
\nobreak

Let us now turn to the $N\to \Delta$ quadrupole 
transition moment. 
This observable and the related $E2/M1$ ratio are exactly zero 
in the symmetric additive quark model \cite{Mor65}. 
The inclusion of tensor forces due to one-gluon 
exchange between quarks leads to small $D$-state admixtures
 $a_{{D}_M}$  ($b_{{D}_S}, \, b_{{D}_M}$)
in the nucleon ($\Delta $) ground states wave functions of eq.(\ref{cfm}) 
and to non-zero $C2$ and $E2$ transition amplitudes \cite{Ger82,Isg82}. 
The magnitude of this configuration
mixing effect is, however, too small. Using the admixture coefficients 
of ref.\cite{Kon80} one obtains a transition
quadrupole moment  $ Q^{imp}_{N\to \Delta}= -0.0022 $ fm$^2$
calculated from the one-body spatial current density. A similar calculation 
using the one-body charge density and the admixture coefficients of 
ref. \cite{Gia90} gives $ Q^{imp}_{N\to \Delta}= -0.0195 $ fm$^2$ 
(for $b=0.613$ fm). In any case, these values are 
much smaller than the empirical  $ Q^{exp}_{N\to \Delta}= -0.0787 $ fm$^2$ 
(see sect. 4.5). 
Here, we show that the major part of the small $C2$ 
transition amplitude is probably due to two-body
pion and gluon exchange charge densities. This is analogous to the
neutron charge radius and quadrupole moment discussed previously. 
Although all quarks in the $N$ and the $\Delta$ are assumed
to be in $ S $-states the system 
can absorb a $C2$ or $E2$ photon by simultaneously flipping
the spin of {\it two quarks} (see fig.5). 
A glance at eq.(\ref{rank2}) shows that the two-body charge operators
can indeed induce such a $\Delta S=2$ transition. 
Using the total charge density of eq.(\ref{totcha}) 
and having replaced the initial state by the nucleon wave function,
we obtain from the $q\to 0$ limit of eq.(\ref{quadff}) 
\be
\label{transquad}
Q_{N\to \Delta}=-{1\over \sqrt{2}} 
b^2{(\delta_g+\delta_{\pi})\over 3 m_q} = 
-{1 \over \sqrt{2}} 
b^2 \left( {M_{\Delta}-M_N \over M_N} \right) 
= { 1 \over \sqrt{2}}  
r^2_n.
\ee
The corresponding numerical results are listed in table 4.

\par
Eq.(\ref{transquad}) relates the transition quadrupole moment 
to the neutron charge radius. As in eq.(\ref{quad}) no model parameter
such as $m_q$ or $b$ appears in the final expression.
We view this result as quite significant. 
It is almost needless to say that the simple result 
of eq.(\ref{transquad}) will be modified in a more complete 
calculation including $D$-waves in the nucleon and $\Delta$.
Nevertheless, we expect that eq.(\ref{transquad}) captures the essential 
physics that makes both observables special and interesting:
both $r_n^2$ and $Q_{N \to \Delta}$ are dominated by nonvalence quark 
degrees of freedom.
Clearly, future experimental results must be carefully 
interpreted; the entire transition quadrupole moment {\it cannot} 
be attributed to the $D$-state admixtures in the $N$ and $\Delta$ ground state 
wave functions. The effect of two-body exchange currents must be 
taken into account, if one wants to isolate the effect of the quark-quark 
potential itself. 
If in a future experiment a $N \to \Delta$ transition quadrupole moment 
of the order $r_n^2/\sqrt{2}$ is confirmed it would most certainly 
be evidence for an important role of nonvalence quark degrees of freedom,
i.e. pion and gluon exchange currents between quarks in this
observable.

\par
\bigskip
\goodbreak
\goodbreak

\subsection{Magnetic moments}
\nobreak

\par
\par
\goodbreak 
\subsubsection{Magnetic moments of the $\Delta$ }
\nobreak

\nobreak
\noindent
The magnetic moments of the $\Delta$ 
are defined as the $q \to 0$ limit
of the magnetic dipole form factor \cite{Gar76}
\be
\label{magff}
F_M({\bf q}^2)= {2 \sqrt{6\pi} M_N \over iq} < J M_J\!\!=\!\!J \, \, T M_T\mid 
         {-i\over 4\pi} \int d\Omega_q 
         [ Y^1(\hat{\bf q}) \times {\bf J} ({\bf q}) ]^1 
         \mid J  M_J\!\!=\!\!J \, \,T M_T>,
\ee 
where ${\bf J}$ is the total current operator in eq.(\ref{totcur}). 
In the additive quark model one obtains 
(in units of nuclear magnetons ($\mu_N ={e\over 2M_N}$) ):
\be
\mu_{\Delta}= 3 \, e_{\Delta}.
\ee 
Including two-body exchange currents, we have
\be
\label{mmdel}
\mu_{\Delta}  = \left ( 3 +{2b^2\over 3}M_N \delta_g(b)
+ {3\delta{\pi}(b) \over 2M_N} 
-{6 \over M_N} \left ( V^{conf}(b)+V^{\sigma}(b) \right ) 
\right ) e_{\Delta}.
\ee
\par
The first term in eq.(\ref{mm}) corresponds to the well-known 
single-quark current result $\mu_{\Delta}= \mu_p e_{\Delta}$. 
The remaining terms express the gluon,
pion, and  scalar exchange current contributions, 
through corresponding 
potential matrix elements.
All contributions to the $\Delta$ magnetic moments are proportional to the
charge of the $\Delta$.  Therefore, the $\Delta^0$ magnetic moment is
predicted to be zero. This is in agreement with the additive 
quark model as well as with a recent lattice calculation \cite{Lei92}.
We list our numerical results in table 5.  
Note the large gluon contribution to the $\Delta^{++} $ magnetic moment,
which gets cancelled
by a similarly large {\it scalar} exchange current
correction\footnote{A vector confinement current has the same sign
as the one-gluon exchange current.}.  
The cancellation of the gluon and 
confinement exchange currents is closely connected to the
cancellation of spin-orbit forces in the 
gluon and confinement potentials \cite{Buc94}. 
For this cancellation it is
essential that the condition $M_N(b)=3m_q=939$ MeV be satisfied and
that the harmonic oscillator parameter $b \approx 1/m_q \approx 0.6 $
fm be consistent with the neutron charge radius of eq.(\ref{rnmec}).  
Note that in the case of the $\Delta$ elastic form factors, 
the dominant isovector pion pair 
and pionic exchange currents  proportional to  $(\b{\tau}_1 \times 
\b{\tau}_2 )_3 $ do not contribute. 
Therefore,  we have to include the next-to-leading order 
isovector pion pair-current in eq.(\ref{ppiv}), which 
is of the same order as the isoscalar pion pair-current.  
We then reproduce the general result \cite{Dil94} 
that the $\Delta$ magnetic moments
and radii are proportional to the charge of the $\Delta$ given in
eq.(\ref{chadel}). 
In ref. \cite{Dil94} it has been argued 
that the pion contribution to the isoscalar nucleon magnetic moment
used by Brown {\it et } al. \cite{Bro80}
induces an intolerably large violation of this proportionality.  
Here, we show that if the isovector and isoscalar 
pion exchange currents are consistently
calculated to the same nonrelativistic order, the proportionality
of the $\Delta$ magnetic moments to the charge of the $\Delta$ 
holds even in the presence of pions. 
\par
Our result does not much deviate from the  
experimental value $\mu_{\Delta^{++}}=5.7 \pm 1.0 $
$\mu_N$\cite{Nef78} and is within 
the experimental range $\mu_{\Delta^{++}}=3.7-7.5 \, \mu_N$
given by the Particle Data Group \cite{PDG94}.
However, it is larger than the most recent experimental value
\cite{Bos91}. It should be mentioned 
that the determination of $\mu_{\Delta^{++}}$ from
$\pi p \to \pi p \gamma$ bremsstrahlung experiments
\cite{Bos91,Nef78} needs theoretical input from $\pi N$ scattering
models with $\Delta$ degrees of freedom. Therefore, the extraction of
the ``bare'' $\Delta^{++}$ magnetic moment from the $\pi p$
bremsstrahlung data has a certain model dependence that should not be
underestimated.  We also mention that quark model
calculations, such as the one presented here 
neglect the coupling of the $\Delta$ to the $\pi
N$ decay channel and thus predict ``bare'' electromagnetic moments.
Our result for the $\Delta^{++}$ magnetic moment agrees
reasonably well with a chiral bag model calculation by Krivoruchenko 
\cite{Ber88}. 

\par
For comparison and later use in the $N \to \Delta$ transition 
moment, we give our results for the nucleon magnetic moments 
(in units of nuclear magnetons ($\mu_N ={e\over 2M_N}$) ):
\bea
\label{mm}
\mu_p  & = &  3 +{b^2\over 3} M_N \delta_g(b)
+ M_N \delta_{\pi}(b) \left ( {1  \over 4 M^2_N } - {b^2\over 3} \right )
-M_N \biggl ( ( {1\over \mu^2} + {1\over 3}b^2 ) \delta_{\pi_{\mu}}(b) - (
\mu \leftrightarrow \Lambda ) \biggr ) \nonumber \\ & & -{6 \over M_N}
\left ( V^{conf}(b)+V^{\sigma}(b) \right ) \nonumber \\
\mu_n &  = & -2 -{b^2\over 9} M_N  \delta_g(b)
+M_N \delta_{\pi}(b) \left ( {1  \over 4 M^2_N } + {b^2\over 3} \right )
+M_N \biggl (( {1\over \mu^2} + {1\over 3}b^2 )
\delta_{\pi_{\mu}}(b) - ( \mu \leftrightarrow \Lambda ) \biggr )
\nonumber \\ & & +{4 \over M_N}
\left ( V^{conf}(b)+V^{\sigma}(b) \right ).
\eea

The first term in eq.(\ref{mm}) corresponds to the well-known 
single-quark current result $\mu_{p}= 3 \, \mu_N$ and $\mu_n =-2 \,\mu_N$.
The other terms are the gluon pair, 
pion pair ($\mu_{\pi q \bar q}$), 
pionic ($\mu_{\gamma \pi \pi}$), and scalar exchange current contributions to 
the magnetic moments.
In addition to the cancellation between gluon and confinement exchange
currents, there is a substantial cancellation between the pion pair 
and pionic exchange currents. Consequently, the overall
exchange current effect is about $5-10\%$ of the impulse approximation
result. We have previously shown that the cancellation 
between the pion pair and pionic currents only occurs if the 
$\delta$-function term in the one-pion exchange potential is 
included \cite{Buc94}. 
Recently, we have calculated the magnetic moments of the
entire baryon octet including the exchange currents of fig.2.
We find that the cancellations between various 
two-body currents also occur for the hyperon magnetic moments
\cite{Wag95} provided that the quark core radius $b \approx 1/ m_q$, which 
is consistent with the value required by the neutron charge radius  
of eq.(\ref{rnmec}).

\par
\par
\goodbreak 
\subsubsection{$N \to \Delta$ transition magnetic moment }
\nobreak

\par
Next we calculate the $N \to \Delta$ magnetic transition moment. 
In contrast to the magnetic moments, where both isoscalar and isovector
exchange currents contribute, only isovector currents 
contribute to the $N\to \Delta$ transition magnetic moment.
Using eq.(\ref{mm}) we can express our result 
for the transition moment as 
\be
\label{dnmm}
\mu_{N\to \Delta}= { 2 \sqrt{2} \over 3} 
\left (\mu_{imp}^p + {1\over 2} \mu_{gq\bar q}^p + 
{3\over 2} \left ( \mu^{IV \,\, p}_{\gamma \pi \pi }  + 
\mu^{IV \, \, p}_{\pi q \bar q } \right )  
+\mu_{\sigma}^p + \mu_{conf}^p
   \right).
\ee
We obtain the numbers in the
last two rows of table 5. 
We see that the total 
transition moment is about  $13\% $ lower than the impulse result.
Again, there are substantial cancellations among the different terms.
In particular, eq.(\ref{dnmm}) shows that there is the same cancellation
between the pion pair and pionic contribution as in the nucleon 
magnetic moments. We point out that our analytic result for the total 
pion exchange current contribution  
to the transition magnetic moment, $\mu^{\pi}_{N \to \Delta}=0.176 \,  \mu_N$,
is somewhat larger than a recent phenomenological estimate, which gives 
$\mu^{\pi}_{p\to \Delta^+} \approx 0.074 \, \mu_N$ 
\cite{Glo96}. 

\par
Finally, we would like to point out that the dominant contribution to the
$N \to \Delta$ transition magnetic moment comes from the single quark 
current, i.e. the first term in eq.(\ref{dnmm}). 
One can show that the impulse contribution is proportional
to the overlap of the orbital symmetric nucleon and $\Delta$ wave functions.
This holds true 
even in the presence of configuration mixing provided that the
$D$-state admixture is small.
Therefore, any model in which this overlap is small,
due to, for example, very different values of $b$ in 
the nucleon and $\Delta$ wave functions will give a very small value of the 
$N \to \Delta$ transition moment.

We close this section by summarizing the main points. It 
has been known for some time that 
baryon  magnetic moments are valence quark 
dominated. Our calculation explicitely shows that 
corrections coming from nonvalence quark degrees of freedom,
such as exchange currents, are important
but rarely exceed $15\%$ of the additive quark model value. 
The problem with the underestimation of the $N \to \Delta$ transition
magnetic moment persists also after inclusion of exchange currents.

\bigskip
\bigskip
\goodbreak 
\noindent
\subsection { Magnetic radii }
\nobreak

\nobreak
\noindent 
Magnetic radii of hadrons measure the extension of the spatial current
distribution. As the charge radii, they are interesting quantities
which are quite sensitive to various model assumptions. 
The magnetic radius is defined as the
slope of the magnetic form factor at zero momentum transfer:
\be
\label{magr}
r^2_M= -{6 \over F_M(0)} {d \over d { {\bf q}}^2 } 
F_{M}({\bf q}^2)\mid_{{\bf q}^2=0},
\ee
Analytic expressions for the magnetic radii of the nucleon, 
including exchange currents, were given previously \cite{Buc94}.  Here,
we list the results for the $\Delta$ magnetic radii:

\bea 
r^2_{\Delta} & = & { e_{\Delta} \over \mu_{\Delta}} 
\Biggl \{ 3 b^2 + {11\over 30} M_N b^4 \delta_g(b) 
+ {3 \over 20 M_N} b^2 \left ( 10 \delta_{\pi}(b) 
                        + {3\over 2} b\delta'_{\pi}(b) \right ) \nonumber \\ 
&    & -{9 \over M_N} b^2  V^{conf}(b)  
- {3\over 2 M_N} b^2  \left (4  V^{\sigma}(b) 
  + b{\partial \over \partial b} V^{\sigma}(b) \right )  
\Biggr \}
 + r_{\gamma q}^2. 
\eea
Likewise, we obtain for $N\to \Delta$
magnetic transition radii:
\bea 
r^2_{N\to \Delta} & = & {2 \sqrt{2} \over \mu_{N\to \Delta} }
\Biggl \{  b^2 
+ {11\over 360} M_N b^4 \delta_g(b) 
- {1 \over 60} M_N b^4 \left ( 10 \delta_{\pi}(b) 
                            + {3\over 2} b\delta'_{\pi}(b) \right )  
+ {1 \over 2} r^{2 \ p}_{\gamma \pi \pi} \mu_p \nonumber \\
&    & - {3 \over M_N} b^2  V^{conf}(b)  
- {1\over   2 M_N} b^2  \left (4  V^{\sigma}(b) 
  + b{\partial \over \partial b} V^{\sigma}(b) \right )  \Biggr \} 
+ r_{\gamma q}^2. 
\eea
Here, $r^{2 \ p}_{\gamma \pi \pi}$ is  the pionic current contribution 
to the proton magnetic radius \cite{Buc94} and 
$\delta'_{\pi}(b)= {\partial \over \partial b} \delta_{\pi}(b) $.

\par
As is clearly seen in table 6, the scalar exchange current cancels the
effect of gluon and pion exchange currents to a large extent.  
Note that a vector-type confinement
potential would have the same sign as the gluon contribution and would
completely spoil the agreement obtained.

\bigskip
\goodbreak
\subsection{ The $\gamma + N \to \Delta$ helicity amplitudes}
\nobreak

\nobreak
\noindent
In this section we consider  the helicity amplitudes for the
transition $ \gamma + N \to \Delta  $.  
The transverse helicity amplitudes are defined as 
\bea
\label{heli}
A_{3/2} &= & -e\sqrt{2\pi /\omega}  < \Delta J_z = 3/2 \mid 
\b{\epsilon} \cdot  {\bf J} \mid N J_z = 1/2 > \nonumber \\
A_{1/2} &= & -e\sqrt{2\pi /\omega}
 < \Delta J_z = 1/2 \mid 
\b{\epsilon} \cdot  {\bf J} \mid N J_z =-1/2 >,
\eea 
where $e^2=1/137$.
In table 7, we show our results for the transverse helicity amplitudes.
For the $ N \to \Delta$ transition, only $ M1 $ and $E2$ multipoles
contribute. 
In this paper we calculate the
$E2$ contribution from the charge density using Siegert's theorem, 
which relates  the transverse electric
multipoles to the Coulomb multipoles in the long wave-length limit. 
It was noted \cite{Dre84,Bor86}
that a calculation of the $E2$ multipole via the charge density is 
to be preferred. In addition to the reasons already mentioned in 
ref.\cite{Dre84,Bor86} there may be an even more important reason
for this large discrepancy between a calculation based on the charge 
and spatial current density. We conjecture that the reason for 
the large difference is that the former includes spatial 
exchange current corrections of spin-orbit type by virtue of 
Siegert's theorem while the latter does not. The issue deserves
further study. In any case, the advantage of using Siegert's theorem 
clearly outweighs the error induced 
by using the long wave-length limit in a situation which involves a 
substantial momentum transfer of $ q \approx M_{\Delta}-M_N $.  
The relation between the multipole form factors and the helicity 
amplitudes is in the center of mass frame \cite{Gia90} 

\bea
\label{heliform}
A_{3/2}({\bf q}^2) &= & -\sqrt{3\pi \omega} \left ( {e \over 2 M_N} \right )  
 \left ( F_{M}^{N \to \Delta}({\bf q}^2) -
{M_N \omega  \over 6} \, F_{Q}^{N \to \Delta}({\bf q}^2) \right ) \nonumber \\
A_{1/2}({\bf q}^2) &= & -\sqrt{\pi \omega} \left ( {e \over 2 M_N} \right )  
 \left (  F_{M}^{N \to \Delta}({\bf q}^2) 
+3 \, {M_N \omega  \over 6} \, F_{Q}^{ N \to \Delta}({\bf q}^2) \right ),
\eea 
where $F_M$ and $F_Q$ 
are the magnetic and quadrupole transition form factors, which are normalized
to the magnetic (sect. 4.4) and quadrupole (sect. 4.2) transition moments.
The relation between our  
$F_{M}^{ N \to \Delta}$ and $F_{Q}^{ N \to \Delta}$ and 
Giannini's \cite{Gia90}  dimensionless $G_{M1}$ and $G_{E2}$ 
is:  $G_{M1}=  ({\sqrt{6}/2}) \, F_{M}^{N \to \Delta}$ 
and $G_{E2}= - ({\omega M_N \sqrt{6} /12}) \, F_{Q}^{N \to \Delta}$. 
The origin of the factors multiplying the magnetic 
and quadrupole form factors in eq.(\ref{heliform}) is 
explained in ref. \cite{Wir87}. 

\par
There have been previous 
calculations  of  two-body current contributions to the $N\to \Delta$ 
transition in the CQM, but not for the $\Delta$ electromagnetic 
moments and radii. 
Ohta \cite{Oht79},  calculates 
two-body currents resulting from 
minimal substitution in the one-gluon exchange and  a scalar 
confinement potential as well as relativistic corrections
to the single-quark current but does not consider pion exchange currents.
In this early calculation, large  
anomalous magnetic moments for the quarks $\kappa=1.83$ were used. 
This  leads to   an unconventional 
nonrelativistic impulse result   
$A_{3/2}(NRI) =-505 \, \,  10^{-3}$ GeV$^{-1/2} $
 which is drastically reduced by large relativistic corrections to the 
single-quark current,  
$A_{3/2}(RCI) =97 \, \,  10^{-3} $ GeV$^{-1/2} $, 
and an even larger  
contribution of the two-body currents,  
$A_{3/2}(EXC) =189 \, \, 10^{-3} $ GeV$^{-1/2} $.
However, large anomalous magnetic moments for the constituent quarks 
are in conflict with general current algebra arguments 
\cite{Wei90} and with explicit calculations 
in the Nambu-Jona-Lasinio model \cite{Vog90}.

\par
Robson \cite{Rob93} has included the pion pair and pionic exchange currents
resulting from minimal substitution in the one-pion exchange 
potential but ignores gluons. 
He finds that the pionic current is small  and neglects this
contribution. In contrast, our calculation shows that the pionic current 
is big and negative. It completely cancels the positive pion pair-current.
The total pion contribution has thus the same sign as the impulse result.
This cancellation between pion pair and pionic currents 
is closely connected with a similar cancellation in the nucleon magnetic 
moments \cite{Buc94} (see also table 5). 
Another difference is that in ref. \cite{Rob93} 
the $E2$ contribution to the helicity amplitudes has been
neglected. However, it is in the $E2$ amplitude where the exchange currents 
are most clearly seen.
\par
Finally, we  give our result for the $ E2/M1$ ratio using 
the definition of Kumano \cite{Ber88}  
\be 
{E2 \over M1 }  =  { 1 \over 3 } \, { A_{1/2}(E2)  \over A_{1/2}(M1) }  
=  {\omega M_N \over 6} \, {Q_{ N \to \Delta} \over \mu_{ N \to \Delta} } =-0.035.
\ee  
Our predicted $E2/M1$ ratio 
is somewhat larger than the recent experimental value extracted from 
photo-pionproduction at MAMI in Mainz which 
gives $(E2/M1)_{exp}= -0.025 \pm 0.002$ \cite{Bec95}. 
The LEGS-BNL data (see, for example, the 
article by D'Angelo in ref.\cite{Eri94}) 
seem to favor a larger  $(E2/M1)_{exp} =-0.03 $.
Comparing with other theoretical predictions,
our result agrees well with 
Skyrme model results, $E2/M1 \simeq -(0.02- 0.05) $ 
\cite{Wir87} and $E2/M1=-0.037$ \cite{Aba96} and dynamical models for
photo-pionproduction $E2/M1=-0.031$ \cite{NBL90}. 
Note that our $E2$ amplitude 
$G_{E2}(0)=-{\omega M_N \sqrt{6} \over 12} Q_{N \to \Delta}= 0.105 $ 
compares reasonably well with the phenomenological analysis of 
Devenish {\it et} al. \cite{Dev76}, which gives 
$G_{E2}(0)=0.02 G_{M1}(0) \approx 0.1$ \cite{Gia90}.
Our prediction is based on the parameter-free
result of eq.(\ref{transquad}) which relates 
the transition quadrupole moment to the neutron charge radius.

Very recently, there has been a new determination of the $E2/M1$ ratio, 
applying the speed plot technique to fixed-$t$ dispersion relations
for the photo-pionproduction amplitudes. The authors include new 
photo-pionproduction  data from the continuous electron beam facilities in
Mainz and Bonn. They obtain $E2/M1= -0.035$ \cite{Han96} in excellent
agreement with our quark model prediction including exchange currents.
\footnote{If we use $\mu_{N \to \Delta} \approx 4 \mu_N$ as  
the empirical value for the transition magnetic moment
we obtain $E2/M1= -0.022$.}
However, several caveats are in order here.
First, the inclusion of configuration mixing would 
certainly modify the numerical value for this ratio. 
Second, we underestimate the empirical 
$N \to \Delta$ transition magnetic
moment. Third, the extraction of a $\gamma N \to \Delta$ 
photocoupling from the photo-pionproduction data is not
model-independent and it would be much safer to calculate the complete 
photo pionproduction multipoles before comparing with experiment \cite{Wil96}.
Nevertheless, our prediction of a {\it large} $E2/M1$ ratio 
emphasizes the important role of nonvalence quark degrees of freedom 
in the transition quadrupole moment,
irrespective of whether one refers to them as meson cloud
of the nucleon, sea-quark degrees of freedom, or exchange currents. 

Next, we show in fig.6 the four-momentum dependence of the 
helicity amplitudes.
We observe that exchange currents contribute to the $A_{3/2}$ amplitude 
between $7\%$ at $Q^2=0$ and $19\%$ of the impulse result 
at $Q^2=0.5$ GeV$^2$. The small value at $Q^2=0$ is firstly due to 
cancellations of different exchange current contributions 
to $F^{N\to \Delta}_{M}$. Secondly, the exchange current
contributions to $F^{N\to \Delta}_{M}$ and $F^{N\to \Delta}_{Q}$ form factors  
interfere destructively in the $A_{3/2}$ amplitude (see also table 7).
On the other hand, in the $A_{1/2}$ amplitude the  
exchange current dominated $F^{N\to \Delta}_{Q}$ form factor enters with
an additional weight factor of three and the exchange current 
contributions to $F^{N\to \Delta}_{M}$
and $F^{N\to \Delta}_{Q}$ form factors interfere constructively.
Therefore, for small momentum transfers the $A_{1/2}$ helicity amplitude is 
appreciably influenced by exchange currents. For example, at 
$Q^2=0$ their contribution is 27$\%$ of the impulse result.

Finally, with respect to the helicity amplitudes of other resonances, we 
have recently calculated their effect for the $M1$ excitation of the
Roper resonance \cite{Buc96}. In this case, the inclusion of exchange 
currents gives for both the proton and neutron an encouraging agreement 
with the empirical values for the $A_{1/2}$ amplitudes. More work is
needed to systematically study the effect of exchange currents 
in the photocouplings of higher resonances. 

\bigskip
\bigskip 
\goodbreak
\section{Summary}
\nobreak

\nobreak
\noindent
In summary, the interaction between quarks manifests itself not 
only in various two-body potentials, but also in corresponding
two-body corrections to the electromagnetic current operator.
These must be included if the total current is to be conserved.
The exchange current operators describe the coupling of the photon to 
nonvalence degrees of freedom (e.g. quark-antiquark pairs) 
not included in the mass, size, and wave
function of the constituent quarks. 
We have pointed out that most
previous calculations of electromagnetic properties built on free
quark currents are incomplete because they violate current conservation
even in lowest order. 

\par
In the present paper we have calculated the electromagnetic radii and moments 
of  the nucleon and the $\Delta$-isobar, as well as the corresponding 
transition radii and moments in the nonrelativistic quark model.
Our main purpose was to study to what extent the
theoretical predictions for these observables 
are modified by the inclusion of the leading order relativistic 
corrections of {\it two-body nature}  in the electromagnetic current.
All observables were calculated with a single set of parameters in order 
to see how the different two-body currents affect various 
observables. Our numerical results clearly show the importance of individual 
exchange current contributions.

\par
With respect to the magnetic moments we found that although individual 
exchange current corrections can be quite large, their overall effect
typically changes the additive quark model result by less than $15 \%$.
This clearly shows that magnetic moments are to a large extent 
valence quark dominated.
In particular, for the $N \to \Delta $ transition magnetic moment, 
a large discrepancy between our theoretical prediction 
and the experimental result is left unexplained even after inclusion
of exchange currents. 

\par
In the case of quadrupole moments, 
we have shown that even if there is no explicit $D$-state admixture
in the $\Delta$ wave function, 
one still obtains a large contribution to the $\Delta$ quadrupole moment
and to the corresponding $N\to \Delta$ transition quadrupole moment
due to two-body pion and gluon exchange currents. 
This is depicted in Fig.5 where an $E2(C2)$ photon can be absorbed 
on a correlated quark pair even if all three quarks in the nucleon are 
in $S$-states. Without two-body exchange currents the $E2(C2)$ amplitude would 
be exactly zero in the present approximation, which neglects configuration 
mixing. Configuration mixing alone is too small to explain the 
empirical $E2(C2)$ amplitude. 
We find that the $E2(C2)$ transition to the $\Delta$ 
is mainly a {\it two-body} process involving the simultaneous spin-flip
of two quarks.

\par
We have, based on the inclusion of exchange currents, derived a number
of analytic relations between the $\Delta -N$  mass splitting and
the electromagnetic observables of the $\Delta-N$ system; in particular, 
our eqs.(\ref{quad},\ref{transquad}) suggest
that the quadrupole moment of the $\Delta$ and the $N \to \Delta$ 
transition quadrupole moment are closely related  
to the neutron charge radius, which, in turn, is related to the
quark core size of the nucleon and the $\Delta - N$ mass difference
according to eq.(\ref{rnmec}).
Because these relations are derived for pure $S$-wave functions
they can only be approximately valid. 
We have previously shown 
\cite{Buc91} that a more consistent 
calculation including both configuration mixing and exchange currents
leads to deviations of some $10-20\%$ between, for example, the prediction 
of eq.(\ref{rnmec}) and the total result including configuration mixing.
Therefore, eqs.(\ref{rnmec},\ref{quad},\ref{transquad}) are very useful. 
First, the numerical estimates, e.g. for the neutron charge radius
are in excellent agreement with experiment, and 
although there is no experimental information on the $\Delta$ 
quadrupole moment, previous and very recent extractions of the 
transition quadrupole moment indicate a value consistent with 
the prediction of eq.(\ref{transquad}).
Second, they seem to correctly describe the underlying physics common to 
these observables: $r_n^2$, $Q_{\Delta}$  and $Q_{N \to \Delta}$  
are almost exclusively dominated by nonvalence 
quark degrees of freedom.
Third, these relations make the underlying connection between 
the excitation spectrum of the nucleon (potentials) 
and electromagnetic properties (two-body currents) explicit.
We cannot resist the temptation to speculate whether  
eqs.({\ref{quad}, \ref{transquad}) are of a somewhat more general 
validity than their derivation would suggest.

Our prediction for the $E2/M1$ ratio, which is  
based on our analytic expressions
for $Q_{N\to \Delta}$ and $\mu_{N \to \Delta} $ results in $E2/M1=-0.035$. 
This value is significantly larger than the value estimated by the Particle 
Data Group \cite{PDG94} but is consistent with a recent reanalysis of
photo-pionproduction data from several experiments \cite{Han96}.
If we use the empirical value for the $\mu_{\Delta}$ we obtain
$E2/M1=-0.022$ in agreement with the recent Mainz experiment \cite{Bec95}. 

\par
Clearly, there are a number of other effects, 
such as configuration mixing, relativistic boost corrections, 
small anomalous magnetic moments 
of the quarks, strangeness content of the nucleon 
and $\Delta$, etc. that should be included in a more detailed analysis.
Nevertheless, it is safe to conclude that 
the residual spin-dependent interactions manifest themselves not
only in excited state admixtures to ground state wave functions
but also in the form of two-body exchange currents between quarks. 
Exchange currents must be included in the theoretical interpretation 
of experimental results before one can draw conclusions about details of 
the quark-quark 
interaction. In particular, we find that the $E2$-amplitude, 
in photo-pionproduction is  predominantly 
a {\it two-quark spin-flip transition}; 
it is to a much lesser extent a consequence of small $D$-states 
in the nucleon. Hence, the experimental confirmation of a large 
$E2$-amplitude in the $N \to \Delta $ transition would be 
evidence for pion and gluon exchange currents between quarks.

\bigskip
\bigskip
\bigskip

\noindent
{\bf Acknowledgement:}
\medskip
\noindent
We thank Andreas Wirzba for useful correspondence. 

\vfill
\eject

\vfill
\eject
%
%
%
%
%
%
\goodbreak
\begin{table}[htb]
\caption[Quark model parameters]{Quark model parameters.
%
%
}
\begin{center} 
\begin{tabular}[h]{|r|r|r|r|r|r|} \hline
$b\,$ [fm] & $\alpha_s$ & $a_c$ [MeV\,fm$^{-2}]$ &
$m_{\sigma}$ [MeV] & $g^2_{\sigma}/( 4 \pi)$ & $ \Lambda$ [fm$^{-1}] $ \\ 
\hline
0.613 & 1.093 & 20.20 & 675 & 0.554 & 4.2 \\
\hline
\end{tabular}
\end{center} 
\end{table}
%
%
%
%
\goodbreak
\begin{table}[htb]
\caption[Nucleon mass]{
Contribution of the kinetic energy (without the rest mass term) 
and individual potential 
terms in the Hamiltonian 
to the nucleon mass of eq.(\ref{massN})
and the gluon ($\delta_{g}$) and pion ($\delta_{\pi}$)
contributions to the $\Delta -N$ mass splitting of eq.(\ref{M-M}).
$cc$: color Coulomb part of $V^{OGEP}$,
$\delta$: $\delta$-function part of $V^{OGEP}$.
All entries are in [MeV]. }
\begin{center} 
\begin{tabular}[h]{|l|l|l|l|l|l|l|l||l|l|} \hline
Term & $ T^{kin}$ & $V^{conf}$ &
$V^{OGEP}_{cc}$ & $V^{OGEP}_{\delta}$ & $V^{\pi}$ &
$V^{\sigma}$ & Total & $\delta_g$ & $\delta_{\pi}$\\
\hline
 & 496.6 & 182.2 & -561.2 & 49.5 & -118.9 & -48.1 & 0.0 & 197.9 & 95.1 \\
\hline
\end{tabular}
\end{center} 
\end{table}
%

%
\begin{table}[htb]
\caption[Charge radii]{Nucleon and $\Delta$(1232)
charge radii including two-body exchange currents;
i: impulse; g: gluon; $\pi$: pion; $\sigma$:
$\sigma$-meson; c: confinement; t: total=impulse + gluon + pion +
sigma + confinement.  A finite electromagnetic quark size $r^2_{\gamma
q}=0.36$ fm$^2$ is used.  The experimental proton and neutron charge
radii are $r_p=0.862 \pm 0.012 $ fm and $\sqrt{\mid r^2_n}\mid
=0.345\pm0.003 $ fm, respectively \cite{Sim80}.  The charge radius of
the $\Delta^{0}$ is zero in the present model.  All entries are in
[fm$^2$] except for the total result which is in [fm].
}
\begin{center}
\nobreak
\begin{tabular}[htb]{| r | r | r | r | r | r | r |} \hline
 & $r^2_i$ & $r^2_{g}$ & $ r^2_{\pi}$ & $ r^2_{\sigma}$ &
$r^2_{c}$ & $ \sqrt{ \mid r^2_{t} \mid} $
\\ \hline \hline
$p $ & 0.736 & $ 0.119 $ & $ -0.057 $ & $0.041$ & $-0.174 $ & $ 0.815
$ \\ \hline
$ n $ & $ 0.000$ & $-0.079 $ & $ -0.038$ & $ 0.000$ & $ 0.000 $ & $
0.342 $ \\ \hline
$\Delta $ & 0.736 & $ 0.198 $ & $ -0.019 $ & $0.041$ & $-0.174 $ & $
0.884 $ \\ \hline \hline
\end{tabular}
\end{center}
\end{table}
%
\begin{table}[htb]
\caption[Quadrupole moments]{$\Delta$(1232) quadrupole moments
and $N \to \Delta$ transition quadrupole moments, including 
two-body exchange currents;
i: impulse; g: gluon; $\pi$: pion; $\sigma$:
$\sigma$-meson; c: confinement; t: total=impulse + gluon + pion +sigma +conf.
As in the neutron charge radius, spin-independent scalar exchange
currents do not contribute to the $\Delta$ quadrupole moments. 
The quadrupole moment of the $\Delta^{0}$ is zero in the present model.  
The experimental values for the transition quadrupole moment are: 
$Q_{N \to \Delta}=-0.0439$ fm$^2$ using the empirical values for 
the helicity amplitudes \cite{PDG94};
$Q_{N \to \Delta}=-0.0787$ fm$^2$ using the phenomenological analysis 
of ref.\cite{Dev76}; a recent Mainz analysis favors an even larger
value $Q_{N \to \Delta}=-0.1109$ fm$^2$ \cite{Han96}.  
All entries are in [fm$^2$].
}
\begin{center}
\nobreak
\begin{tabular}[htb]{| r | r | r | r | r | r | r |} \hline
 & $Q_i$ & $Q_{g}$ & $ Q_{\pi}$ & $ Q_{\sigma}$ & $Q_{c}$ & $ Q_{t}  $ \\ 
\hline \hline
$ \Delta^{++} $ & $ 0.000$ & $-0.158 $ & $ -0.0761$ & $ 0.000$ & $ 0.000 $ & $
-0.234 $ \\ \hline
$\Delta^{+} $ & 0.000 & $ -0.079 $ & $ -0.038 $ & $0.000$ & $ 0.000 $ & 
$ -0.117$ \\ \hline
$ \Delta^{0} $ & 0.000 & $ 0.000 $ & $ 0.000 $ & $0.000$ & $0.000 $ & $
0.000 $ \\  \hline
$\Delta^{-} $ & 0.000 & $ 0.079 $ & $ 0.038 $ & $0.000$ & $ 0.000 $ & 
$ 0.117$ \\ \hline \hline
$p\to\Delta^{+} $ & 0.000 & $ -0.056 $ & $ -0.027 $ & $0.000$ & $ 0.000 $ & 
$ -0.083$ \\ \hline
$ n\to \Delta^{0} $ & 0.000 & $ -0.056 $ & $ -0.027 $ & $0.000$ & $0.000 $ & $
-0.083 $ \\ \hline \hline
\end{tabular}
\end{center}
\end{table}
%
%
\goodbreak
\begin{table}[htb]
\caption[Magnetic moments]{Nucleon and $\Delta$(1232) magnetic moments,
and $N\to \Delta$ transition magnetic moments 
including two-body exchange currents;
i: impulse; g: gluon; $\pi$: pion;
$\sigma$: $\sigma$-meson; c: confinement; t: total=impulse + gluon +
pion + sigma + confinement.  
The contribution of the pion pair 
($ \pi q \bar q$) and the pionic ($\gamma \pi \pi $) currents 
are separately listed.  
The experimental proton and neutron
magnetic moments are $\mu_p=2.792847386(63) $ $\mu_N$ and 
$\mu_n=-1.91304275(45)$ $\mu_N$,
respectively \cite{PDG94}. The  experimental range for the
$\Delta^{++}$ magnetic moment is $\mu_{\Delta^{++}}=3.7-7.5$ $\mu_N$ 
\cite{PDG94}. An older value is $\mu_{\Delta^{++}}=5.7 \pm 1.0\,$
$\mu_N$\cite{Nef78} while the most recent value is 
$\mu_{\Delta^{++}}=4.52 \pm 0.50\, \mu_N $ \cite{Bos91}. 
The experimental value for the $N \to \Delta$ transition magnetic moment
is $\mu_{p \to \Delta^+} \approx 4.0 \mu_N$ \cite{Gia90}. 
 All entries are in $\mu_N$.  }
\begin{center}
\nobreak
\begin{tabular}[h]{| l | r | r | r | r | r | r | r |} \hline
 & $\b{\mu}_i$ & $\b{\mu}_{g}$ & $\b{\mu}_{\pi q\bar q}$ &
$\b{\mu}_{\gamma \pi \pi}$ & $\b{\mu}_{\sigma}$ & $\b{\mu}_c$ &
$\b{\mu}_{t} $ \\ \hline \hline
$p $ & $3.000 $ & $ 0.598 $ & $ -0.262 $ & $ 0.411$ & $0.308$ &
$-1.164 $ & $ 2.890 $ \\ \hline
$ n $ & $ -2.000$ & $-0.199 $ & $0.313 $ & $-0.411$ & $-0.205$ & $
0.776 $ & $-1.726 $ \\ \hline \hline
$ \Delta^{++} $ & $ 6.000 $ & $2.391$ & $0.304 $ & $0.000 $ & $ 0.615
$ & $-2.328$ & $6.981 $
\\ \hline 
$ \Delta^{+} $ & $ 3.000 $ & $1.195$ & $0.152 $ & $0.000 $ & $ 0.308 $
& $-1.164$ & $3.491 $
\\ \hline 
$ \Delta^{0} $ & $ 0.000 $ & $0.000 $ & $0.000 $ & $0.000 $ & $ 0.000
$ & $ 0.000$ & $ 0.000 $
\\ \hline 
$ \Delta^{-} $ & $ -3.000 $ & $-1.195$ & $-0.152 $ & $0.000 $ & $-0.308
$ & $1.164$ & $-3.491 $
\\ \hline \hline
$ p\to \Delta^{+} $ & $ 2.828 $ & $0.282$ & $-0.406 $ & $0.582 $ & $ 0.290 $
& $-1.098$ & $2.477 $ \\ \hline 
$ n\to \Delta^{0} $ & $ 2.828 $ & $0.282$ & $-0.406 $ & $0.582 $ & $ 0.290 $
& $-1.098$ & $2.477 $ \\ \hline \hline 
\end{tabular} 
\end{center}
\end{table}

%
%
%
\begin{table}[htb]
\caption[Magnetic radii] {
Magnetic radii of the nucleon and $\Delta$(1232) including two-body
exchange currents;
i: impulse; g: gluon; $\pi$: pion; $\sigma$:
$\sigma$-meson; c: confinement; t: total=impulse + gluon + pion +
sigma + confinement.  
The contribution of 
the pion pair ($ \pi q \bar q$) current and the pionic current 
($\gamma \pi \pi $) are listed separately.  
The magnetic radius of the $\Delta^0$ is zero.
A finite electromagnetic quark size, $r^2_{\gamma q}=0.36$ fm$^2$, is used. 
The experimental proton and neutron magnetic
radii are $r^2_p=0.858 \pm 0.056$ fm and $r^2_n=0.876 \pm 0.070$ 
fm \cite{Sim80}.  All entries are in
[fm$^2$], except for total results which are in [fm].
%
}
\begin{center}
\nobreak
\begin{tabular}[h]{| r | r | r | r | r | r | r | r |} \hline
 & $r^2_i$ & $r^2_{g}$ & $ r^2_{\pi q \bar q}$ & $
r^2_{\gamma \pi \pi }$ & $ r^2_{\sigma}$ & $r^2_{c}$ & $ \sqrt{ \mid
r^2_{t} \mid} $ \\ \hline
$p $ & $0.764 $ & $ 0.117 $ & $-0.053 $ & $0.185 $ & $0.058 $ &
$-0.372 $ & $0.836 $ \\ \hline
$ n $ & $ 0.852 $ & $0.065 $ & $-0.105 $ & $0.309 $ & $0.065 $ & $
-0.415 $ & $0.878$ \\ \hline
$ \Delta $ & $ 0.632 $ & $0.194$ & $0.025 $ & $0.000 $ & $0.048$ &
$ -0.308 $ & $0.769 $ \\ \hline \hline
$ p\to \Delta^{+} $ & $ 0.840  $ &  
 $0.064$ & $-0.096 $  & $0.305 $ & $0.064 $ & $ -0.409 $
&  $ 0.876 $ \\ \hline 
\end{tabular} 
\end{center}
\end{table}

\begin{table}[htb]
\caption[Helicity amplitudes]{
The $A_{3/2}$ and $A_{1/2}$ helicity amplitudes 
for the process $\gamma + N\to \Delta(1232)$, 
including two-body exchange currents; i: impulse; g: gluon; 
$\pi q {\bar q} $: pion pair; 
$\gamma \pi \pi  $: pionic; 
$\sigma$: $\sigma$-meson; c: confinement; 
t: total=impulse + gluon + pion +sigma +conf.
The $M1$ and $E2$ parts of the helicity amplitudes as well as their sum 
are listed. 
The experimental helicity amplitudes are $A_{3/2}=-257 \pm 8$ and
$A_{1/2}=-141 \pm 5$ \cite{PDG94}. 
A previous  analysis gave $A_{1/2}=-84 \pm 5$  \cite{Tan85}.  
Our results are given at ${\bf q}^2=0$. All entries are given in
standard units of [$10^{-3}$ GeV$^{-1/2}$].
}
\begin{center}
\nobreak
\begin{tabular}[h]{| r | r | r | r | r | r | r | r | r |} \hline
  & $A^{}_i$ & $A^{}_{g}$ & $ A^{}_{\pi q\bar q}$ 
& $ A^{}_{\gamma \pi \pi}$ 
& $ A^{}_{\sigma}$ & 
$A^{}_{c}$ & $ A^{}_{t}  $ & $ A^{}_{exp}  $ \\ \hline \hline
%
%
%
%
$A^{3/2}_{p\to\Delta^{+}}(M1) $ &- 200.7 & $-20.0  $ & $ 28.8 $ & $ -41.3 $ 
& $-20.6$ & $ 77.9 $ & $ -175.8 $  & $ -253.8 $ \\ \hline
$A^{3/2}_{p\to\Delta^{+}}(E2) $ & 0.0 & $ -4.1  $ & $ -2.0 $ & $ 0.0 $ 
& $ 0.0$ & $ 0.0 $ & $ -6.1 $  & $ -3.2 $ \\ \hline
$A^{3/2}_{p\to\Delta^{+}}(T) $ &- 200.7 & $-24.1  $ & $ 26.8 $ & $ -41.3 $ 
& $-20.6$  & $ 77.9 $ & $ -181.9 $  & $ -257.0 $ \\ \hline \hline
%
%
%
%
$A^{1/2}_{ p\to \Delta^{+}}(M1)$ & -115.9 & $ -11.5 $ & $ 16.7 $  & $ -23.8 $ 
& $ -11.9$ & $ 45.0 $ & $-101.5$  & $-146.5$  \\ \hline
$A^{1/2}_{ p\to \Delta^{+}}(E2)$ &  0.0 & $7.1  $ & $ 3.4 $  & $ 0.0 $ 
& $ 0.0 $ & $ 0.0  $ & $ 10.5 $  & $ 5.5 $  \\ \hline
$A^{1/2}_{ p\to \Delta^{+}}(T)$ & -115.9 & $-4.4  $ & $ 20.1 $  & $ -23.8 $ 
& $ -11.9$ & $ 45.0 $ & $-90.9$  & $-141.0$  \\ \hline
\end{tabular}
\end{center}
\end{table}
\goodbreak

\vskip 25 truecm
\vfill
\eject
\newpage

\goodbreak

\begin{figure} 

\centerline {\bf FIGURE CAPTIONS} 
\bigskip

\begin{itemize}

\item[Fig.1]{Residual (a) one-gluon,
(b) one-pion, and (c) one-sigma exchange potentials between
constituent quarks. The hadronic size $r_q$ of the constituent quarks
is indicated by small dots.}

\bigskip

%

\item[Fig.2]{One-body and two-body exchange currents 
between quarks: (a) impulse, (b) gluon pair, (c) pion pair, (d)
pionic, (e) scalar pair. The finite electromagnetic size of the
constituent quarks and the pion is indicated by the filled circles.  }

\bigskip
%
\item[Fig.3]{Pion loop contribution to the electromagnetic form factor of the 
constituent quarks. Vector meson dominance relates the finite 
electromagnetic size of the constituent quarks to the vector meson mass 
$r^2_{\gamma q}\approx 6/m_{\rho}^2$  \cite{Vog90}. 
}

\bigskip


\item[Fig.4]{One-body and two-body contributions to the 
quadrupole moment of 
the $\Delta$ and to the $N\to \Delta$ transition quadrupole moment.
In diagram (a) the photon is absorbed on a single quark which 
remains in a positive energy state after 
the absorption of the photon. The dominant contribution of this 
diagram is obtained by sandwiching the standard {\it one-body} 
current between baryon wave functions. 
These wave functions must contain a tensor force induced 
$D$-state in order that the system can 
absorb a $C2$ or $E2$ photon through a {\it single-quark} transition.
In diagram (b) the photon couples to a quark-antiquark pair in the baryon and
the system can absorb a $C2$ or $E2$ photon on {\it two quarks}, 
even if  all quarks are in $S$-states.
This contribution is effectively described by the {\it two-body} exchange 
charge operators. Similar diagrams can be drawn for pion-exchange between 
quarks. }

\bigskip

\item[Fig.5]{Pion and gluon exchange current contribution to the 
$E2$ transition form factor. 
A major part of the $E2$ transition form factor is due to
photon absorption on a correlated
pair of quarks, interacting via gluon and  pion exchange. 
The $E2$ photon simultaneously flips the spin of {\it two quarks}
in the nucleon leading to the $\Delta(1232)$. This process is more
important than the one where an $E2$ photon is absorbed on a 
single-quark moving in a $D$-wave (see fig. 4a).}

\item[Fig.6]
{The $A_{1/2}(Q^2)$ (a) 
and $A_{3/2}(Q^2)$ (b)  helicity amplitudes as a function of the 
four-momentum transfer $Q$. 
Here, we keep $\omega_{cm}=258$ MeV fixed and vary 
the three-momentum transfer ${\bf q}$.
The individual two-body exchange current contributions are shown separately.
}
\end{itemize}
\end{figure}

\begin{figure}[htb]
\label{Fig.1}
$$\hspace{0.2cm} \mbox{
\epsfxsize 12.0 true cm
\epsfysize 6.0 true cm
\setbox0= \vbox{
\hbox {
\epsfbox{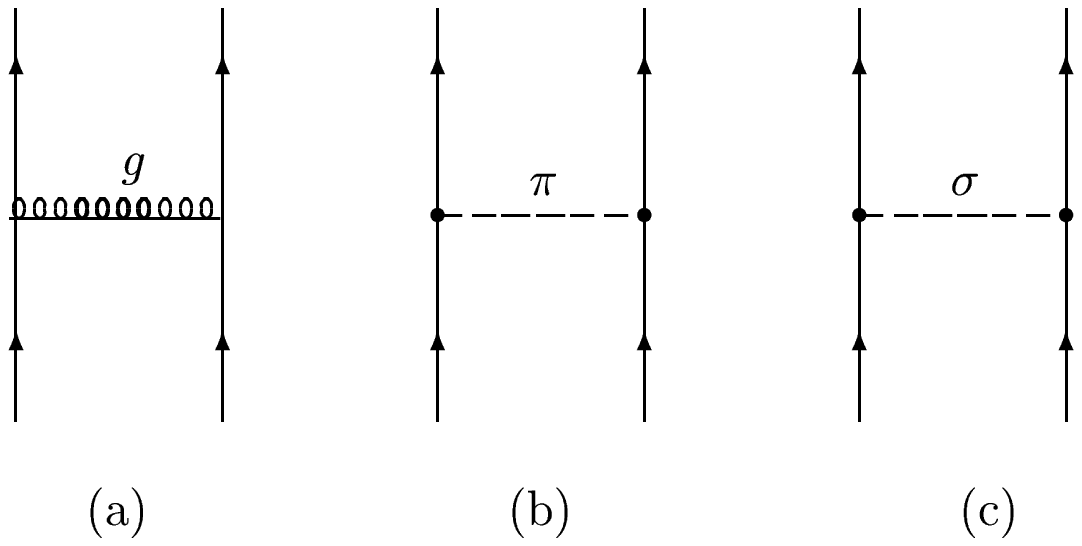}
} 
} 
\box0
} $$
\vspace{1.0cm}
\caption[Residual interactions]{  }
\end{figure}

\vspace{2.0 cm}
\begin{figure}[htb]
\label{Fig.2}
$$\hspace{-0.75cm} \mbox{
\epsfxsize 16.0 true cm
\epsfysize 6.0 true cm
\setbox0= \vbox{
\hbox {
\epsfbox{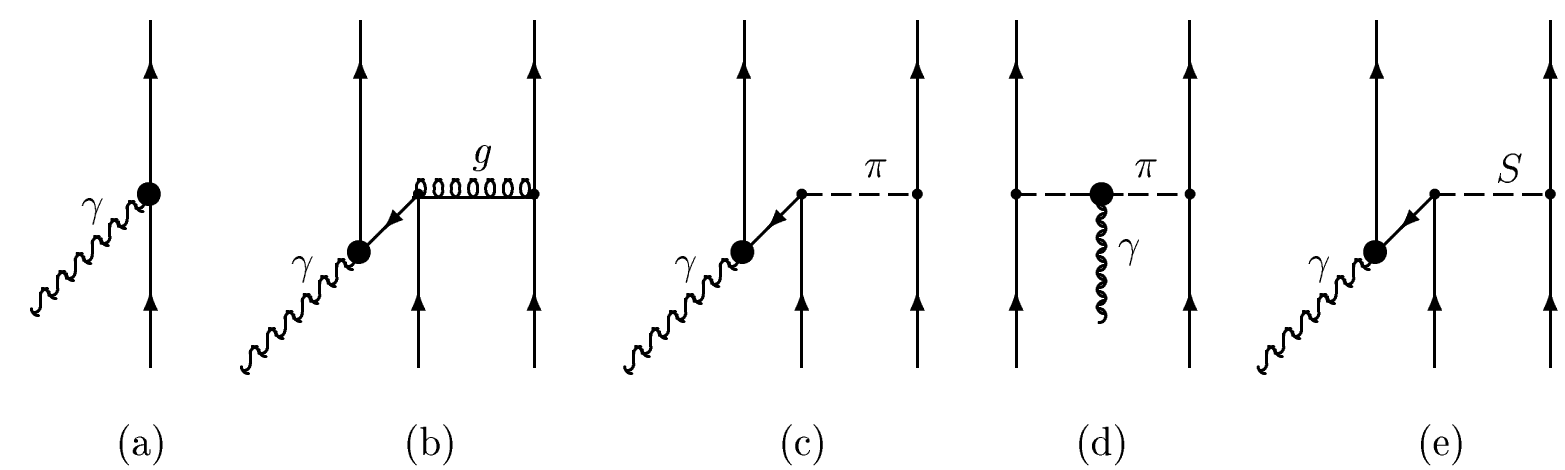}
} 
} 
\box0
} $$
\vspace{-1.0cm}
\caption[Austauschstr\"ome]{   }
\end{figure}

\begin{figure}[htb]
\label{Fig.3}
$$\hspace{0.2cm} \mbox{
\epsfxsize 13.5 true cm
\epsfysize 10.0 true cm
\setbox0= \vbox{
\hbox {
\epsfbox{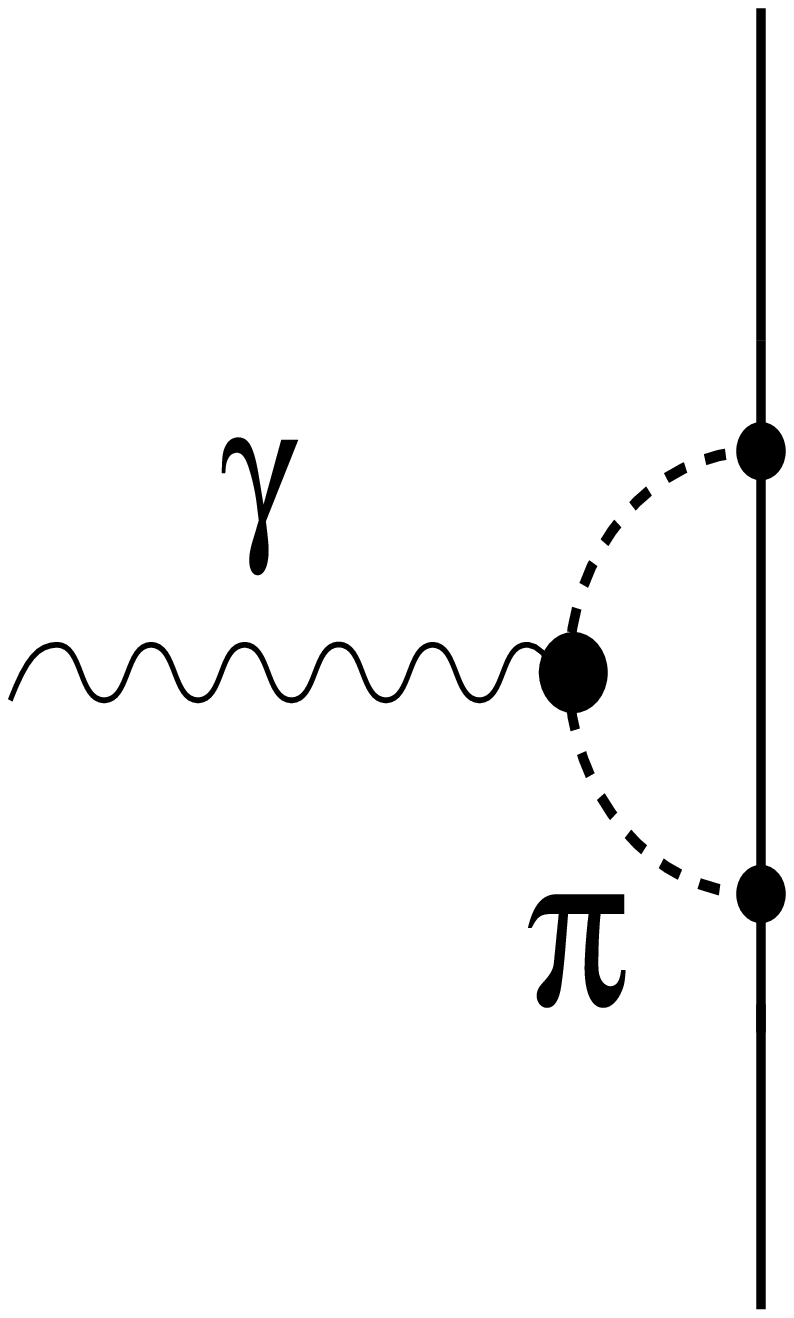}
} 
} 
\box0
} $$
\vspace{-2.0cm}
\caption[VMD]{   }
\end{figure}

\vspace{2.0 cm}
\begin{figure}[htb]
\label{Fig.4}
$$\hspace{0.2cm} \mbox{
\epsfxsize 10.0 true cm
\epsfysize 5.75 true cm
\setbox0= \vbox{
\hbox {
\epsfbox{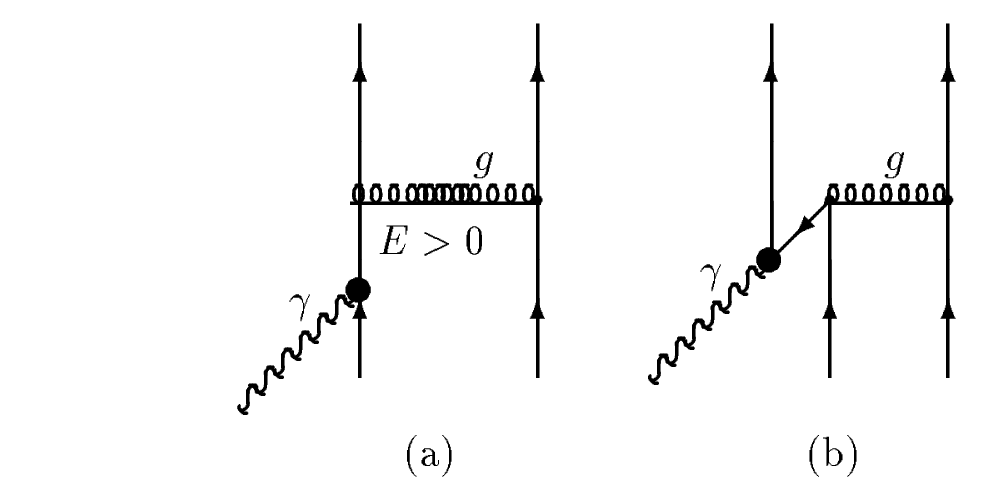}
} 
} 
\box0
} $$
\vspace{0.3cm}
\caption[Interpretation]{    }
\end{figure}


\begin{figure}[htb]
\label{Fig.5}
$$\hspace{0.2cm} \mbox{
\epsfxsize 15.2 true cm
\epsfysize 6.0 true cm
\setbox0= \vbox{
\hbox {
\epsfbox{fig5.ps}
} 
} 
\box0
} $$
\vspace{-1.5cm}
\caption{    }
\end{figure}

\vspace{1.0 cm}
\begin{figure}[htb]
\label{Fig.6}
$$\hspace{0.2cm} \mbox{
\epsfxsize 15.0 true cm
\epsfysize 10.0 true cm
\setbox0= \vbox{
\hbox {
\epsfbox{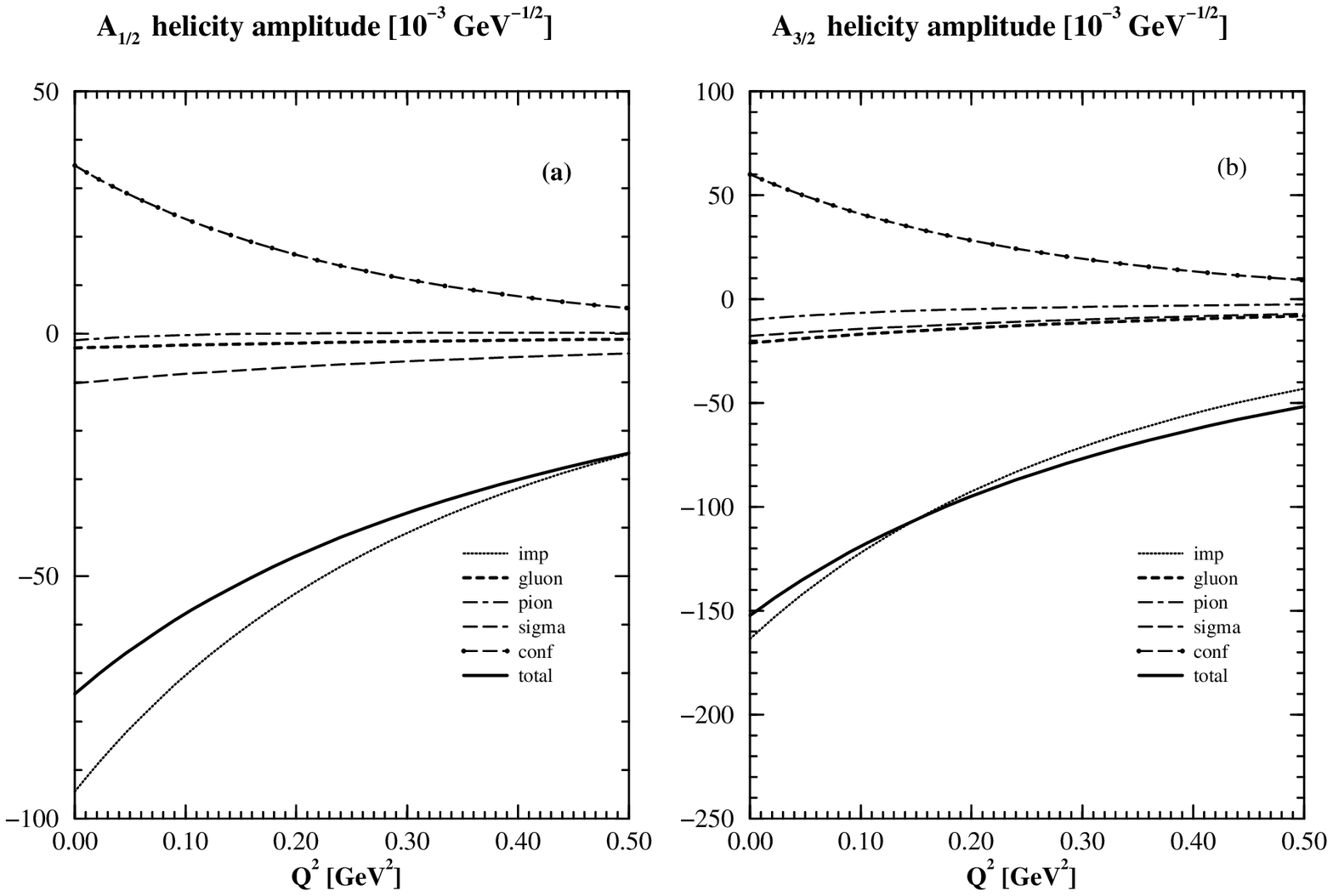}
} 
} 
\box0
} $$
\vspace{-1.5cm}
\caption{     }
\end{figure}

\end{document}